\documentclass[twocolumn]{aastex63}
\usepackage{comment}
\usepackage{amsmath}	
\usepackage[normalem]{ulem}
\usepackage{amssymb}
\usepackage{footnote}
\usepackage{appendix}
\usepackage[graphicx]{realboxes}
\usepackage{blindtext}
\usepackage{rotating}
\usepackage{cleveref}
\usepackage{threeparttable}
\usepackage{tablefootnote}
\usepackage{multirow}
\usepackage[super]{nth}

\graphicspath{{./}{figures/}}

\newcommand{\dmunits}{\ensuremath{{\rm pc \, cm^{-3}}}}
\newcommand{\dmcosmic}{\ensuremath{{\rm DM}_{\rm cosmic}}}
\newcommand{\dmacosmic}{\ensuremath{\langle {\rm DM}_{\rm cosmic} \rangle}}
\newcommand{\dmfrb}{\ensuremath{{\rm DM}_{\rm FRB}}}
\newcommand{\dmigm}{\ensuremath{{\rm DM}_{\rm IGM}}}
\newcommand{\dmhost}{\ensuremath{{\rm DM}_{\rm host}}}

\newcommand{\dmhalo}{\ensuremath{{\rm DM}_{\rm halo}}}

\newcommand{\dmhaloeg}{\ensuremath{{\rm DM}_{\rm halo, EG}}}

\newcommand{\dmism}{\ensuremath{{\rm DM}_{\rm ISM}}}
\newcommand{\dmeg}{\ensuremath{{\rm DM}_{\rm EG}}}

\newcommand{\zdm}{\ensuremath{z-{\rm DM}}}
\newcommand{\zdmeg}{$z$-DM$_{\rm EG}$}

\newcommand{\pdmz}{\ensuremath{p({\rm DM}|z)}}

\newcommand{\pdmegz}{\ensuremath{p({\rm DM_{\rm EG}}|z)}}

\newcommand{\sigdm}{\sigma_{\rm DM}}

\newcommand{\lhost}{\ensuremath{{\mu}_{\rm host}}}
\newcommand{\lsigma}{\ensuremath{{\sigma}_{\rm host}}}

\newcommand{\hunits}{\ensuremath{{\rm km \, s^{-1} \, Mpc^{-1}}}}

\newcommand{\icshigh}{CRAFT/ICS \ensuremath{1.6\,{\rm GHz}}}
\newcommand{\icsmid}{CRAFT/ICS \ensuremath{1.3\,{\rm GHz}}}
\newcommand{\icslow}{CRAFT/ICS \ensuremath{900\,{\rm MHz}}}

\newcommand{\NFRB}{78 }
\newcommand{\NFRBz}{21 }
\newcommand{\NFRBnoZ}{57 }

\shorttitle{Variance of the Macquart Relation}
\shortauthors{Baptista et al.}

\begin{document}

\sloppy

\title{Measuring the Variance of the Macquart Relation in z-DM Modeling}

\correspondingauthor{J. Baptista}
\email{jaymarie@stanford.edu}

\newcommand{\Yale}{\affiliation{Department of Astronomy and Astrophysics, Yale University, New Haven, CT 06520, USA}}

\newcommand{\UCSC}{\affiliation{Department of Astronomy and Astrophysics, University of California, Santa Cruz, CA 95064, USA}}

\newcommand{\NU}{\affiliation{Center for Interdisciplinary Exploration and Research in Astrophysics (CIERA) and Department of Physics and Astronomy, Northwestern University, Evanston, IL 60208, USA}}

\newcommand{\IPMU}{\affiliation{Kavli Institute for the Physics and Mathematics of the Universe (Kavli IPMU), 5-1-5 Kashiwanoha, Kashiwa, 277-8583, Japan}}

\newcommand{\NAOJ}{\affiliation{Division of Science, National Astronomical Observatory of Japan, 2-21-1 Osawa, Mitaka, Tokyo 181-8588, Japan}}

\newcommand{\STSCI}{\affiliation{
Space Telescope Science Institute, Baltimore, MD 21218, USA
}}

\newcommand{\JHU}{\affiliation{Department of Physics and Astronomy, Johns Hopkins University, Baltimore, MD 21218, USA}}

\newcommand{\iceland}{\affiliation{Centre for Astrophysics and Cosmology, Science Institute, University of Iceland, Dunhagi 5, 107 Reykjav\'ik, Iceland}
}

\newcommand{\Macquarie}{\affiliation{Department of Physics \& Astronomy, Macquarie University, NSW 2109, Australia}
}

\newcommand{\astrophotonics}{\affiliation{Astronomy, Astrophysics and Astrophotonics Research Centre, Macquarie University, Sydney, NSW 2109, Australia}
}

\newcommand{\CSIRO}{\affiliation{Australia Telescope National Facility, CSIRO Astronomy and Space Science, PO Box 76, Epping, NSW 1710, Australia}
}

\newcommand{\Swinburne}{\affiliation{Centre for Astrophysics and Supercomputing, Swinburne University of Technology, Hawthorn, VIC 3122, Australia}}

\newcommand{\tata}{\affiliation{Department of Astronomy and Astrophysics, Tata Institute of Fundamental Research, Mumbai, 400005, India}}

\newcommand{\NCRA}{\affiliation{National Centre for Radio Astrophysics, Post Bag 3, Ganeshkhind, Pune, 411007, India}}

\newcommand{\PUCV}{\affiliation{Instituto de F\'isica, Pontificia Universidad Cat\'olica de Valpara\'iso, Casilla 4059, Valpara\'iso, Chile}}

\newcommand{\MPI}{\affiliation{Max Planck Institute for Astrophysics, Karl-Schwarzschild-Str. 1, 85741 Garching, Germany}}

\newcommand{\card}{\affiliation{Cardiff Hub for for Astrophysics Research and Technology, School of Physics and Astronomy, Cardiff University, Queen’s Buildings, The Parade, Cardiff CF24 3AA, UK}}

\newcommand{\Curtin}{\affiliation{International Centre for Radio Astronomy Research, Curtin University, Bentley, WA 6102, Australia}}

\newcommand{\Stanford}{\affiliation{Department of Physics, Stanford University, Stanford, CA 94305, USA}}

\newcommand{\KIPAC}{\affiliation{Kavli Institute for Particle Astrophysics & Cosmology, Stanford University, P.O. Box 2450, Stanford, CA 94305, USA}}
\author[0000-0002-9306-1704]{Jay Baptista}
\Yale
\affiliation{Kavli Institute for Particle Astrophysics \& Cosmology, Stanford University, P.O. Box 2450, Stanford, CA 94305, USA}

\author[0000-0002-7738-6875]{J.~Xavier~Prochaska}
\UCSC
\IPMU
\NAOJ

\author{Alexandra G. Mannings}
\UCSC

\author[0000-0002-6437-6176]{C.W. James}
\Curtin


\author[0000-0002-7285-6348]{R.~M.~Shannon}
\Swinburne

\author{Stuart D. Ryder}
\Macquarie

\author{A.~T.~Deller}
\Swinburne

\author{Danica R. Scott}
\Curtin

\author{Marcin Glowacki}
\Curtin

\author{Nicolas Tejos}
\PUCV

\newcommand{\forecastedHubble}{\ensuremath{H_0 = 69.17_{-4.91}^{+5.46}}}
\newcommand{\forecastedHubblewithPrior}{\ensuremath{H_0 = 67.57_{-3.36}^{+3.51}}}

\newcommand{\FnoPrior}{\ensuremath{-0.75_{-0.25}^{+0.33}}}
\newcommand{\FwPrior}{\ensuremath{-0.48^{+0.26}_{-0.18}}}
\newcommand{\FCMB}{\ensuremath{-0.35_{-0.15}^{+0.23}}}
\newcommand{\FSNe}{\ensuremath{-0.52_{-0.17}^{+0.26}}}

\newcommand{\fctFnoPrior}{\ensuremath{-0.58_{-0.15}^{+0.15}}}
\newcommand{\fctFwPrior}{\ensuremath{-0.60_{-0.10}^{+0.09}}}
\newcommand{\fctFCMB}{\ensuremath{-0.52_{-0.06}^{+0.07}}}
\newcommand{\fctFSNe}{\ensuremath{-0.67_{-0.06}^{+0.06}}}

\newcommand{\Hubble}{\ensuremath{85.3_{-8.1}^{+9.4}}}
\newcommand{\HwPrior}{-}

\newcommand{\fctH}{\ensuremath{69.2_{-4.9}^{+5.5}}}
\newcommand{\fctHwPrior}{\ensuremath{67.6_{-3.4}^{+3.5}}}
\newcommand{\Flower}{\ensuremath{\log_{10} F > -0.89}}

\begin{abstract}

The Macquart relation describes the correlation between the 
dispersion measure (DM) of fast radio bursts (FRBs) and the redshift $z$ 
of their host galaxies. The scatter of the Macquart relation 
is sensitive to the distribution of baryons in 
the intergalactic medium (IGM) including those 
ejected from galactic halos through feedback processes. 
The width of the distribution in DMs from the cosmic web
(\dmcosmic) is parameterized by a fluctuation parameter $F$, which 
is related to the cosmic DM variance by 
$\sigdm = F z^{-0.5}$.
In this work, we present a new measurement of $F$ 
using \NFRB FRBs of which \NFRBz have been localized to host galaxies. 
Our analysis simultaneously fits for the Hubble constant $H_0$ and 
the DM distribution due to the FRB host galaxy. 
We find that the fluctuation parameter is degenerate with these parameters, most notably $H_0$, and use a uniform prior on $H_0$ to measure \Flower\, at the $3\sigma$ confidence interval and a 
new constraint on the Hubble constant $H_0 = \Hubble \, \hunits$.
Using a synthetic sample of 100 localized FRBs, the constraint on the fluctuation parameter is improved by a factor of $\sim 2$. Comparing our $F$ measurement to simulated predictions from 
cosmological simulation (IllustrisTNG), we find agreement between $0.4 < z < 2$. However, at $z < 0.4$, the simulations underpredict $F$ which we attribute to the rapidly changing extragalactic DM excess distribution at low redshift. 

\end{abstract}


\section{Introduction} \label{sec:intro}


In galaxy formation models, AGN and stellar feedback have provided mechanisms for regulating star formation and evacuating gas out of low-mass halo \citep{cen2006, dave11}. In simulations without baryonic outflows, galaxies simply produce too many stars and have higher than observed star formation rates \citep{dave11}. These outflow processes are also critical in understanding how the IGM becomes enriched and how the galaxies and the IGM co-evolve. 

Not only is understanding the nature of feedback crucial in reproducing realistic galaxy properties in cosmological-baryonic simulations but also in understanding the channels where these ``missing" baryons may have left halos and are prevented from re-accretion. Gas accretion onto galaxies from cold gas filaments is exceptionally efficient. Baryonic feedback is a preventative process that not only enriches the IGM but also removes baryons and restricts accretion from the IGM \citep{keres05}. For example, in the \textsc{Simba} suite of cosmological hydrodynamic simulations, feedback from AGN jets can cause 80\% of baryons in halos to be evacuated by $z=0$ \citep{dave19, appleby21, sorini22}.

A comparison of simulation suites shows that different feedback prescriptions can eject baryons at various distances beyond the halo boundary—feeding gas into the reservoir of diffuse baryons \citep[e.g.][]{ayromlou23}. Thus, to constrain the strength of AGN and stellar feedback processes, one must be able to constrain the distribution of baryons in the IGM to discriminate between these feedback models.

Determining the distribution of these ejected baryons is difficult. Emission and absorption lines from baryons in the IGM are extremely difficult to detect due to their high temperatures and low densities \citep{fukugita98, cen2006, shull12, mcquinn14}. However, the advent of fast radio bursts (FRBs) presents a new opportunity to probe the intergalactic distribution of baryons and provide a novel approach to measuring feedback strength \citep{mcquinn14, munoz18}.


FRBs are sensitive to the line-of-sight free electron density where the integrated free electron density yields the dispersion measure (DM) of the signal \citep{lorimer07}. A redshift can be estimated if the spatial localization of the FRB overlaps with a galaxy with a known redshift \citep[assuming the FRB progenitor indeed lies within that galaxy;][]{aggarwal21b}. 
This FRB redshift-extragalactic dispersion measure (z-$\dmeg$) correlation known as the Macquart relation \citep{mq2020} is sensitive to cosmological properties of the universe \citep[e.g.][]{j22a}. By using a sophisticated FRB observational model that can account for observational biases, intergalactic gas distribution, burst width, and DM, it is possible to use FRB surveys 
to infer the distribution of baryons in the universe
\citep{mcquinn14, j22a, lee22}. 

Halos with weaker feedback retain their baryons more effectively leading to halos with higher \dmeg\, contributions and voids with lower \dmeg\, contributions \citep{mcquinn14}. An FRB can travel either through extremely low DM voids or pass through extremely high DM halos suddenly, leading to enhanced scatter in the \zdmeg\, distribution. On the other hand, halos with stronger feedback will cause the \zdmeg\, distribution to show less scatter. 
Feedback processes are able to more effectively relocate halo baryons into the IGM, causing inter-halo voids to have higher DM. This leads to a more homogeneous universe. 
The goal of this work is to forecast and measure a precise galactic feedback prescription as determined by a sample of FRBs.

This paper is organized as follows: Section \ref{methods} outlines the modeling of the \zdm\, distribution and how the fluctuation parameter $F$ influences observations of the FRB distribution; Section \ref{ll_survey} presents our measurements on the fluctuation parameter $F$ and the fits on other parameters used in the \zdmeg\, model; Section \ref{ll_forecast} details our forecast on $F$ by sampling 100 synthetic FRBs and finding the probability distribution functions of our model parameters based on the synthetic survey; Section \ref{sec:disc} discusses our results in the context of constraining cosmic feedback strength, parameter degeneracies, and compares our measurements to simulations.


\section{Methods} \label{methods}

\subsection{Basic Formalism}
\label{zdm_model}

This work makes use of the FRB code \textsc{zdm} developed by \cite{j22a} to model observables of FRB populations.
The model assumes that the DM measurement of an FRB can be decomposed as $\dmfrb =  \dmism + \dmhalo + \dmeg$,
where \dmism\, and \dmhalo\, are the DM contributions due to the Milky Way's interstellar medium \citep[modeled using NE2001;][]{ne2001} and diffuse ionized gas in the Galaxy's halo \citep{xyz19, cook23, ravi23}.
The $\dmeg$ term is the extragalactic DM contribution and is decomposed into $\dmeg = \dmcosmic + \dmhost$ where $\dmcosmic = \dmigm + \dmhaloeg$ is the contribution due to baryons in the IGM and intersecting halos in the line-of-sight, and \dmhost\, is the contribution due to the host galaxy of the FRB signal. The contribution from the host (\dmhost) is modeled as a log-normal distribution with a width of $\exp(\mu)$ and a logarithmic scatter of \lsigma\, where $\mu$ and $\lsigma$ are free parameters of the model.


The width of the \zdmeg\, distribution at a fixed redshift is characterized
in part
by the probability distribution of measuring the $\dmcosmic$ of an FRB above or below $\dmacosmic$. This distribution is described by $p_{\rm cosmic}(\Delta)$, where $\Delta \equiv \dmcosmic / \dmacosmic$:

\begin{equation}
    p_{\rm cosmic}(\Delta) = A \Delta^{-\beta} \exp(- \frac{(\Delta^{-\alpha} - C_0)^2}{2 \alpha^2 \sigdm^2}),
\end{equation}
\noindent
where $\alpha \simeq 3$ and $\beta \simeq 3$ are the inner and outer slopes of the gas profile density of intervening halos \citep[based on numerical simulations from][]{mq2020}, $C_0$ shifts the distribution such that $\langle \Delta \rangle = 1$, and $\sigdm$ represents the spread of the distribution \citep{mq2020}. This non-Gaussian probability distribution function is motivated by theoretical treatments of the IGM and galaxy halos \citep{mq2020}. For example, in the limit where $\sigdm$ is small, the distribution becomes Gaussian to capture the Gaussianity of large-scale structures. This is physically motivated as halo gas is more diffuse in this limit and thus contributions to the variance due to halo gas are insignificant. In the limit where $\sigdm$ is large, the halo gas contribution becomes significant \citep{mq2020}.

\begin{figure}
    \centering
    \includegraphics[width=3.5in]{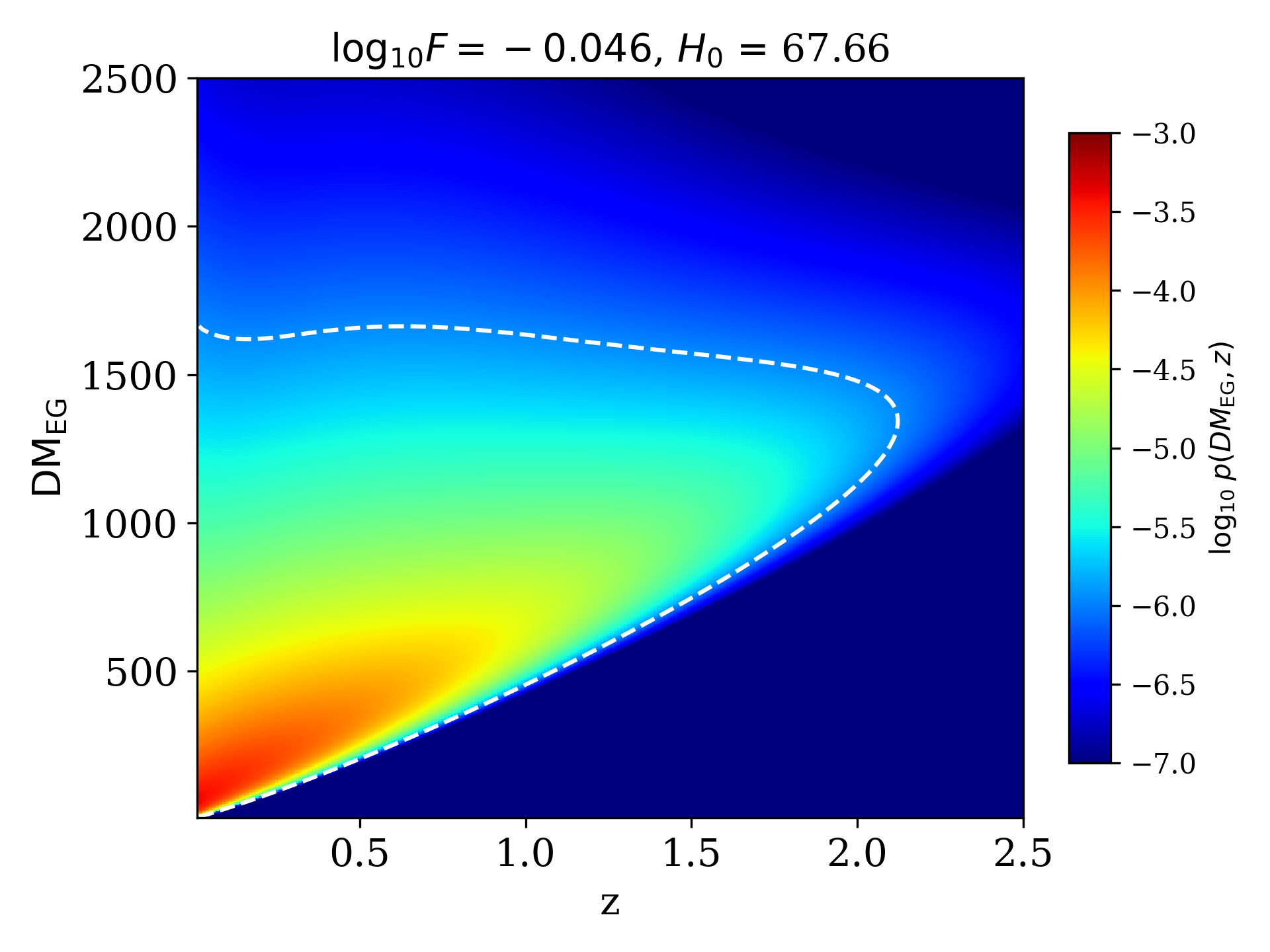}
    \includegraphics[width=3.5in]{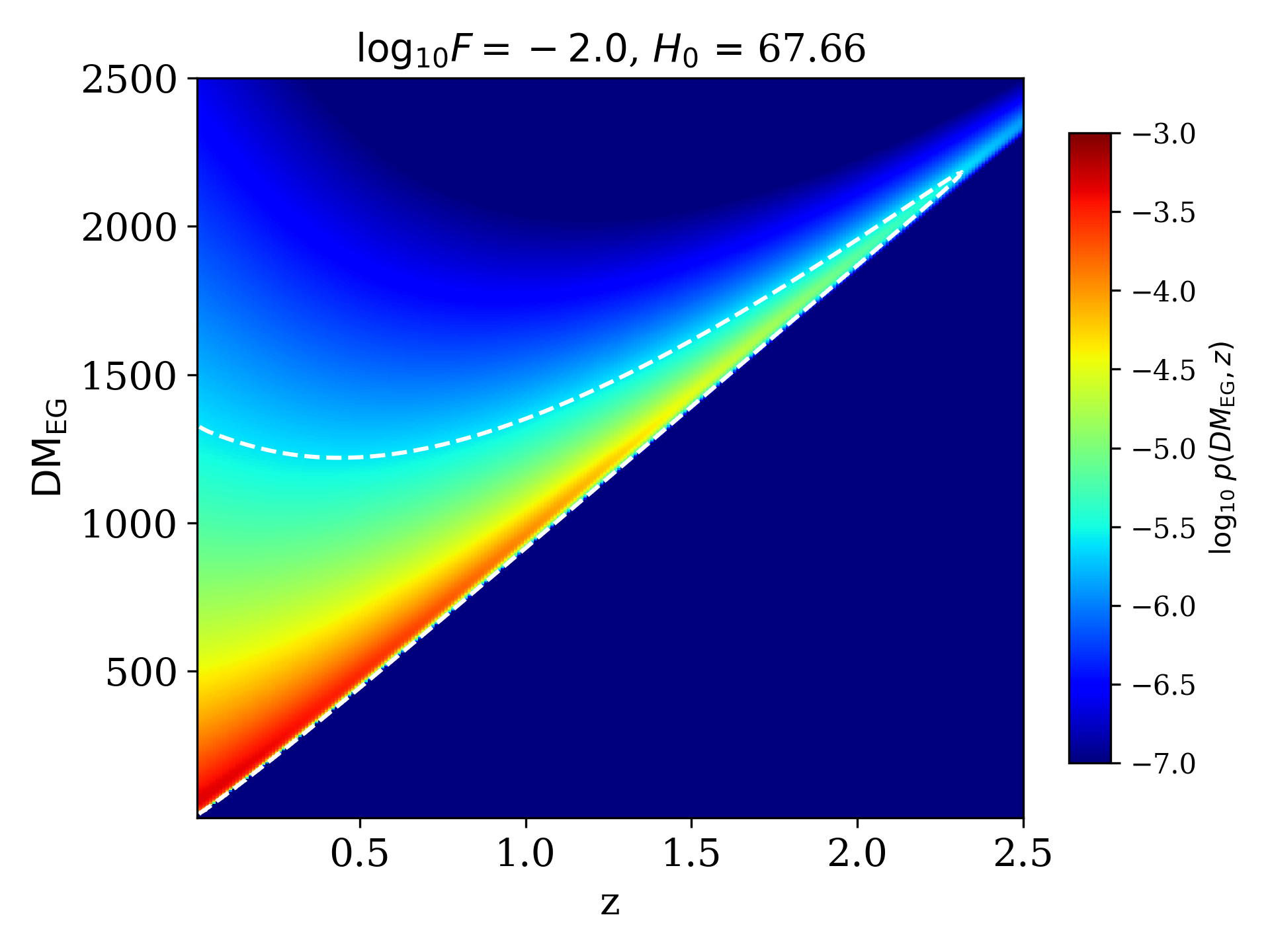}
    \caption{\textbf{Upper panel}: The \pdmegz\, distribution which admits a high fluctuation parameter (low galactic feedback efficiency). \textbf{Lower panel}: The \pdmegz\, distribution which admits a low fluctuation parameter (high galactic feedback efficiency). 
    The white dashed-line indicates the \nth{95} percentile contour. Note that the distribution primarily falls below the mean due to the rare population of high DM FRBs that result from intersections with the host galaxy and/or very massive galaxy halos 
    along the LOS.}
    \label{fig:pzdm_F_compare}
\end{figure}

Owing to the approximately Poisson nature of intersecting
halos, one expects $\sigdm \propto z^{-1/2}$ \citep{mq2020}
and one is motivated to introduce a
fluctuation parameter $F$ 
\begin{equation} \label{eq:sigma_dm}
    \sigdm(\Delta) = Fz^{-0.5} \;\;\; ,
\end{equation}
\noindent
As the fluctuation parameter increases, i.e. $F \sim 1$, the spread of $\dmcosmic$ increases. 
Figure~\ref{fig:pzdm_F_compare} shows \pdmz\ for
two extreme values of $F$ and the resultant, substantial
changes to the width of the \dmeg\ distribution at any
given redshift.

The variance in $\dmeg$, however, is influenced 
by both $\lsigma$ and $F$. However, at high redshift, the contribution to the variance of $\dmeg$ due to $\lsigma$ 
may decrease relative to the contributions by $\dmcosmic$ and, inherently, $F$. In \citet{j22b}, their work assumes that uncertainties attributed to the fixed value of $F$ can be aggregated into uncertainties in $\lsigma$; however, at high redshift, the assumption breaks down as the uncertainty in $F$ becomes larger than the true constraint in $\lsigma$.

Although \citet{mq2020} restrict their fitting of the fluctuation parameter to $F \in [0.09, 0.32]$ based on semi-analytic models, 
we sample a wide range of $F \in [0, 1]$. 
We opt for a logarithmic sampling of the fluctuation parameter
to efficiently sample this domain: $\log_{10} F \in [-2, 0]$.

The additional parameters used in the model include $H_0$ (acceleration of the Universe's expansion), the \dmeg\, contribution due to the FRB host galaxy, and other parameters that govern the FRB luminosity function and redshift distribution. The model assumes that the \dmeg\, contribution from the FRB host galaxy can be modeled as a log-normal distribution with a mean of $\mu_{\rm host}$ (or $\dmhost$) and a spread of $\sigma_{\rm host}$.

In terms of the luminosity function, the maximum burst energy is given as $E_{\rm max}$, and the integral slope of the FRB luminosity function is controlled by $\gamma$. The volumetric burst rate ($\Phi$) is controlled by the parameter $n_{\rm sfr}$ assuming a star-formation rate: $\Phi \propto {\rm SFR(z)}^{n_{\rm sfr}}$. Additionally, $\alpha$ is the spectral index that sets a frequency-dependent FRB rate as $\Phi(z, \nu) = \Phi(z) \nu^{\alpha}$ \citep{j22b}.


\subsection{Measuring $F$ Using FRB Survey Data} \label{survey_data}

To measure the fluctuation parameter, we perform a simultaneous fit of the parameters in the \textsc{zdm} model implemented by \citet{j22b}. We obtain the probability distributions of each parameter by a brute-force grid search based on the ranges specified in Table \ref{tab:fullcube}, and calculating the likelihoods for each permutation of parameter values.

We fit these parameters using both the FRB sample used in \citet{j22b} and newly detected or analyzed
FRBs (see Table \ref{tab:frbs}) which were collected from the Parkes and ASKAP telescopes. Of this sample of \NFRB measured FRBs, \NFRBnoZ FRBs do not have measured redshifts. Constraining the redshift of an FRB greatly increases the statistical power as a single FRB with a redshift can have the same constraining power as roughly 20 FRBs without redshifts \citep{j22b}. Thus our inclusion of seven new CRAFT/ICS FRBs detected over three frequency ranges (four with host redshifts), and our identification of the host galaxy of FRB20211203C 
at $z=0.344$, provides a significant increase in statistical precision.

There is a slight bias in this sample, as we include FRB20220610A, which has an energy exceeding the previously estimated turnover $E_{\rm max}$ by a factor of 3.5--10, depending on the assumed spectral behavior \citep{ryder22}.  FRB20210912A has a lower DM of 1234.5\,\dmunits, but it does not have an identified redshift, perhaps due to the distance to its host galaxy (Marnoch et al., in prep). Therefore, the inclusion of some data is redshift-dependent. Given that our sample is statistically limited, we assume the resulting bias to be small compared to the gain in precision.

In contrast to the \citet{j22b} analysis, we hold model parameters that are not degenerate with $F$ to their fiducial values. These parameters were determined to be non-degenerate with $F$ running the model using a low-resolution grid search on synthetic data to determine if $F$ correlated with any of the other model parameters. From this preliminary analysis, we fix the following parameters that were found to be non-degenerate with $F$: $\alpha$, $\gamma$, $E_{\rm max}$, and $n_{\rm sfr}$. On the other hand, we expect the fluctuation parameter to be degenerate with the other model parameters. 
In particular, we expect the Hubble constant $H_0$--—the cosmological parameter that quantifies the expansion of the universe--—to be degenerate with the fluctuation parameter $F$.

\begin{figure}
    \centering
    \includegraphics[width=3.5in]{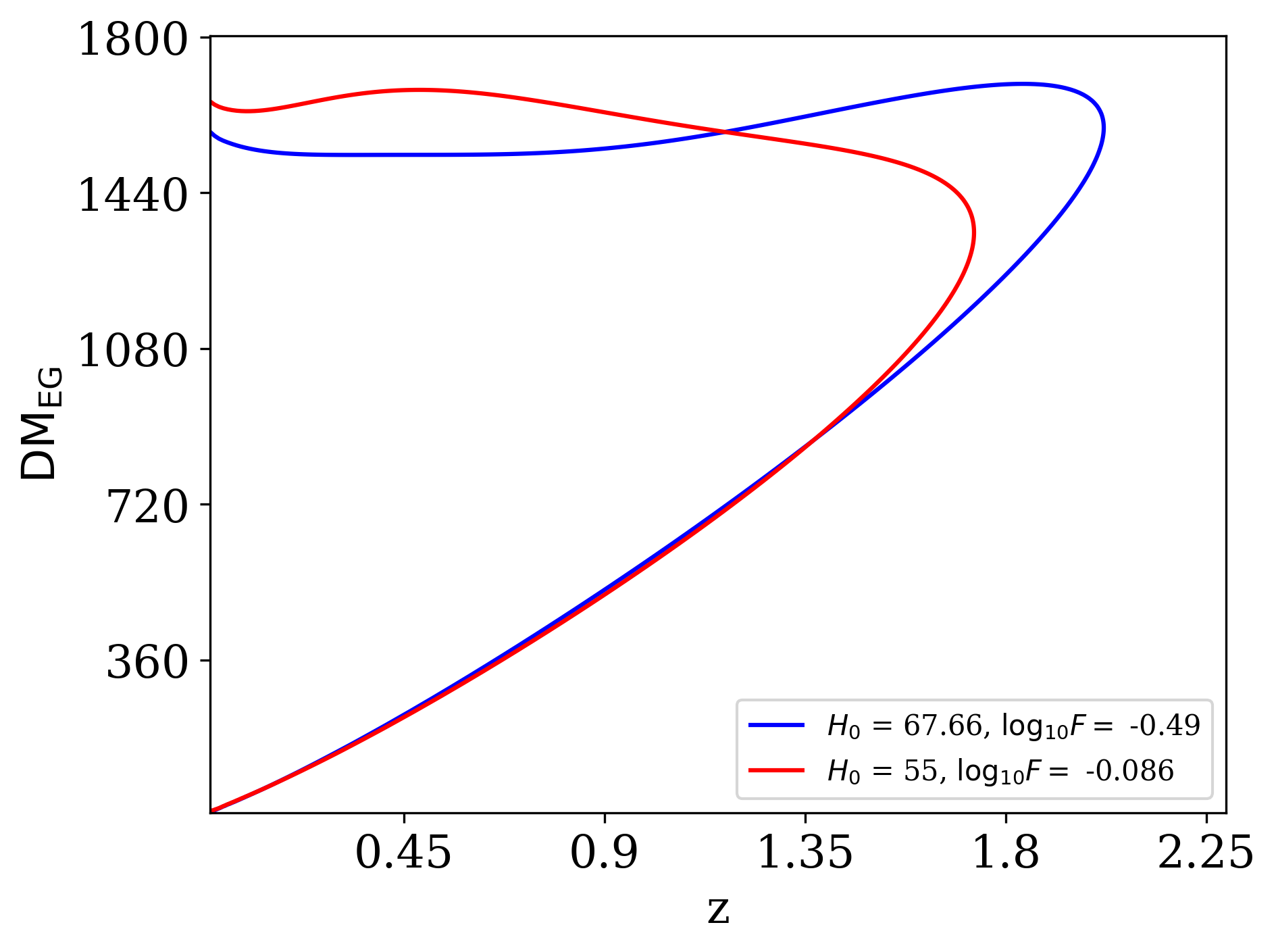}
    \caption{$95^{\rm th}$ percentile contours of two \pdmegz\, distributions with different prescriptions on the Hubble constant $H_0$ and the fluctuation parameter $F$. Note that the lower contours (``the DM cliff") of both models are nearly identical to each other. Since the DM cliff places a stronger constraint on $H_0$ and $F$ than the upper contour, we expect a high degree
    of degeneracy between $H_0$ and $F$. The units of \dmeg\, are in pc cm$^{-3}$ and the units of $H_0$ are $\rm km \, s^{-1} \, Mpc^{-1}$.}
    \label{fig:pzdm_degenerate}
\end{figure}

We examine this degeneracy further in Figure~\ref{fig:pzdm_degenerate} which shows
the \nth{95} percentile contours for 
\pdmz\ for two models with very different 
$F$ and $H_0$ values.
One notes that 
the lower contours 
($\dmeg \lesssim 720, \, z \lesssim 1.3$) of both realizations look nearly identical. Although the contours differ above the mean, the bulk of the constraining power on $F$ is in the lower contour or ``DM cliff".
Therefore, we anticipate $F$ and $H_0$ to be highly correlated.

The distribution at the low DM end of \pdmegz\, exhibits a sharp cut-off and provides strong constraints on $H_0$ since there is a minimum imparted \dmcosmic\, from voids and is not impeded by the \dmeg\, contributions from large-scale structures like filaments or halos.
And while the contours do have modest differences
at high $z$, high \dmeg, these can be difficult
to distinguish from host galaxy contributions
to \dmeg.

\begin{table*}
\centering
\begin{minipage}{170mm} 
\centering
\caption{New FRB detections detected in 2022 used in addition to the FRB surveys used in \citet{j22b}. 
The FRB name, SNR-maximizing DM, \dmism\ estimated using the NE2001 model of \citet{CordesLazio01}, central frequency of observation $\nu$, measured signal-to-noise ratio SNR, redshift $z$, and original reference. Where redshifts are not given, this is because (a): no voltage data were dumped, preventing radio localization; (b) optical follow-up observations are not yet complete; (c) Substantial Galactic extinction has challenged follow-up optical observations; (d) the host galaxy appears too distant to accurately measure a redshift. \label{tab:frbs}
}
\begin{tabular}{ccccccl}
\hline 
Name & DM & \dmism & $\nu$ & SNR & $z$ & Ref.\ \\
& (\dmunits) & (\dmunits) & (MHz) &  & &
\\ 
\hline 

\multicolumn{7}{c}{\icslow} \\
\hline
20211203C & 636.2    & 63.4  & 920.5 & 14.2 & 0.344 & \multirow{3}{*}{Shannon et al. (in prep.)}\\
20220501C & 449.5    & 30.6    & 863.5 & 16.1 & 0.381 &  \\
20220725A & 290.4  & 30.7  & 920.5 & 12.7 & 0.1926 &  \\
\hline 

\multicolumn{7}{c}{\icsmid} \\
\hline
20220531A & 727.0    & 70.0    & \multirow{3}{*}{1271.5} & 9.7 & -- & Shannon et al. (in prep.) \\
20220610A & 1458.1  & 31.0  & & 29.8 & 1.016 & \citet{ryder22} \\
20220918A & 656.8  & 40.7  & & 26.4 & -- &  Shannon et al. (in prep.)\\
\hline 

\multicolumn{7}{c}{\icshigh} \\
\hline
20220105A & 583.0    & 22.0    & 1632.5 & 9.8 & 0.2785 & \multirow{2}{*}{Shannon et al. (in prep.)} \\
20221106A & 344.0  & 34.8  & 1631.5 & 35.1 & -- &  \\
\hline

\end{tabular} 
\end{minipage} 
\end{table*}

\subsection{Forecasting the fluctuation parameter $F$ using Synthetic FRBs} \label{forecasting}

\begin{deluxetable}{cccccc}
\tablewidth{20pt}
\tablecaption{z-DM grid parameters \label{tab:fullcube}}
\tabletypesize{\normalsize}
\tablehead{
\colhead{Parameter} & 
\colhead{Unit} & 
\colhead{Fiducial} &
\colhead{Min} &
\colhead{Max} &
\colhead{N}
}
\startdata 
$H_0$ & km s$^{-1}$ Mpc$^{-1}$ & 67.4 & 60.0 & 80.0 & 21 \\ 
$\log F$ & - & 0.32 & -1.7 & 0 & 30 \\
$\lhost$ & $\log {\rm pc \, cm^{-3}}$ & 2.16 & 1.7 & 2.5 & 10 \\
$\lsigma$ & $\log {\rm pc \, cm^{-3}}$ & 0.51 & 0.2 & 0.9 & 10 \\
$\log_{10} E_{\rm max}$ & $\log {\rm erg}$ & 41.84 & -- & -- & -- \\
$n_{\rm sfr}$ & - & 1.77 & -- & -- & --\\
$\alpha$ & - & 1.54 & -- & -- &--\\
$\gamma$ & - & -1.16 &-- & -- & --\\
\hline 
\enddata 
\tablecomments{This table indicates the parameters of the high-resolution grid run. Non-degenerate parameters are held to the fiducial values. $N$ is the number of cells between the minimum and maximum parameter values.}

\end{deluxetable}

Future radio surveys are expected to widely increase the number of sub-arcsecond
localized FRBs. Thus, the constraining power on $F$ will greatly increase. To explore this scenario we generate a forecast on the fluctuation parameter by replicating our analysis using a synthetic FRB survey. A sample of 100 localized synthetic FRBs was drawn assuming the distribution of FRBs followed the fiducial \zdm\ distribution 
(Table~\ref{tab:fullcube}). With this synthetic survey, we calculate the associated 4D likelihood matrix and make a forecast on the fluctuation parameter by adopting different priors on $H_0$.

\section{Results} \label{sec:res}

\subsection{Parameter Likelihoods from FRB Surveys}
\label{ll_survey}

\begin{figure*}
    \centering
    \includegraphics[width=2.5in]{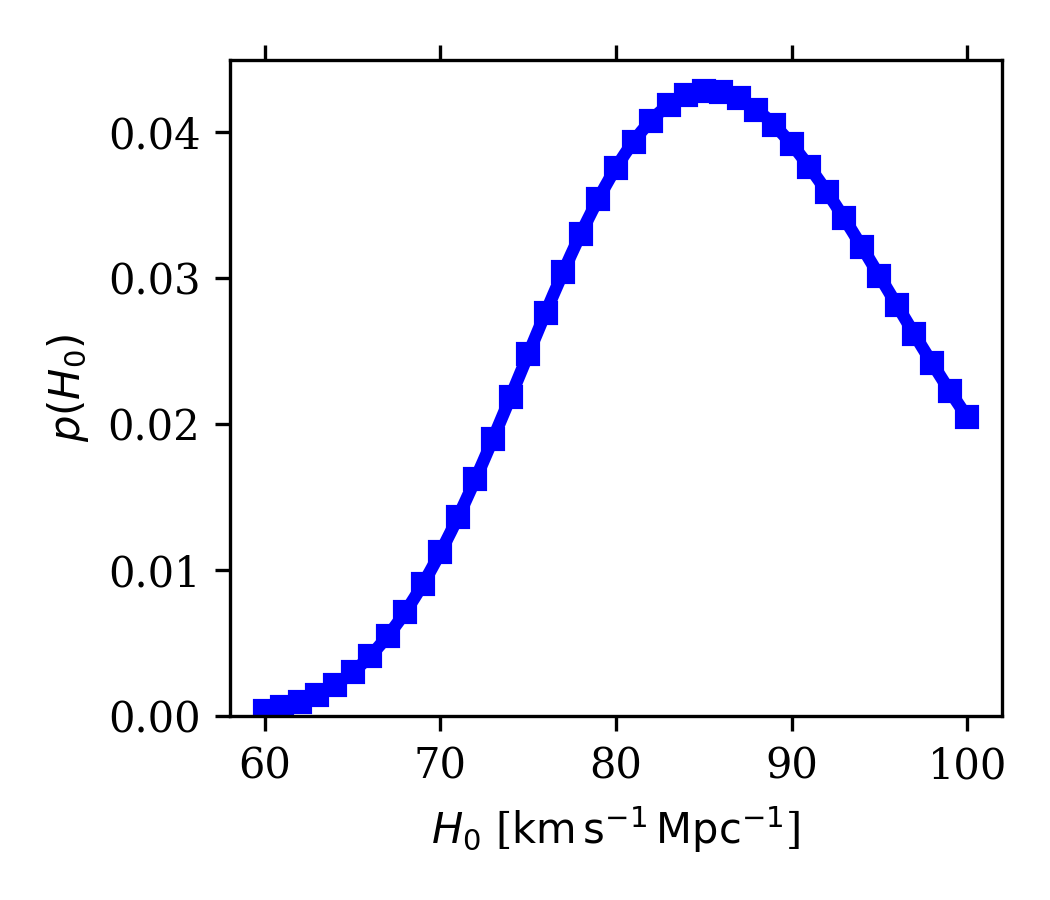} 
    \includegraphics[width=2.5in]{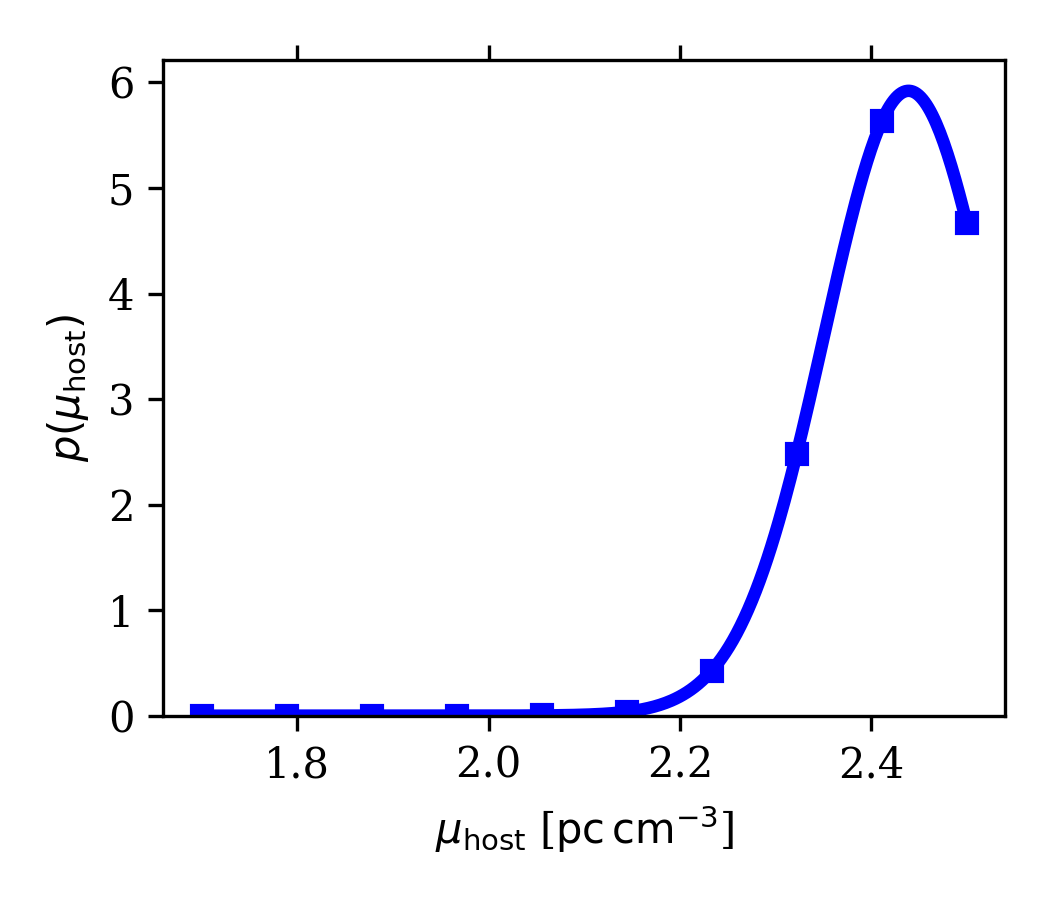} \\
    \includegraphics[width=2.5in]{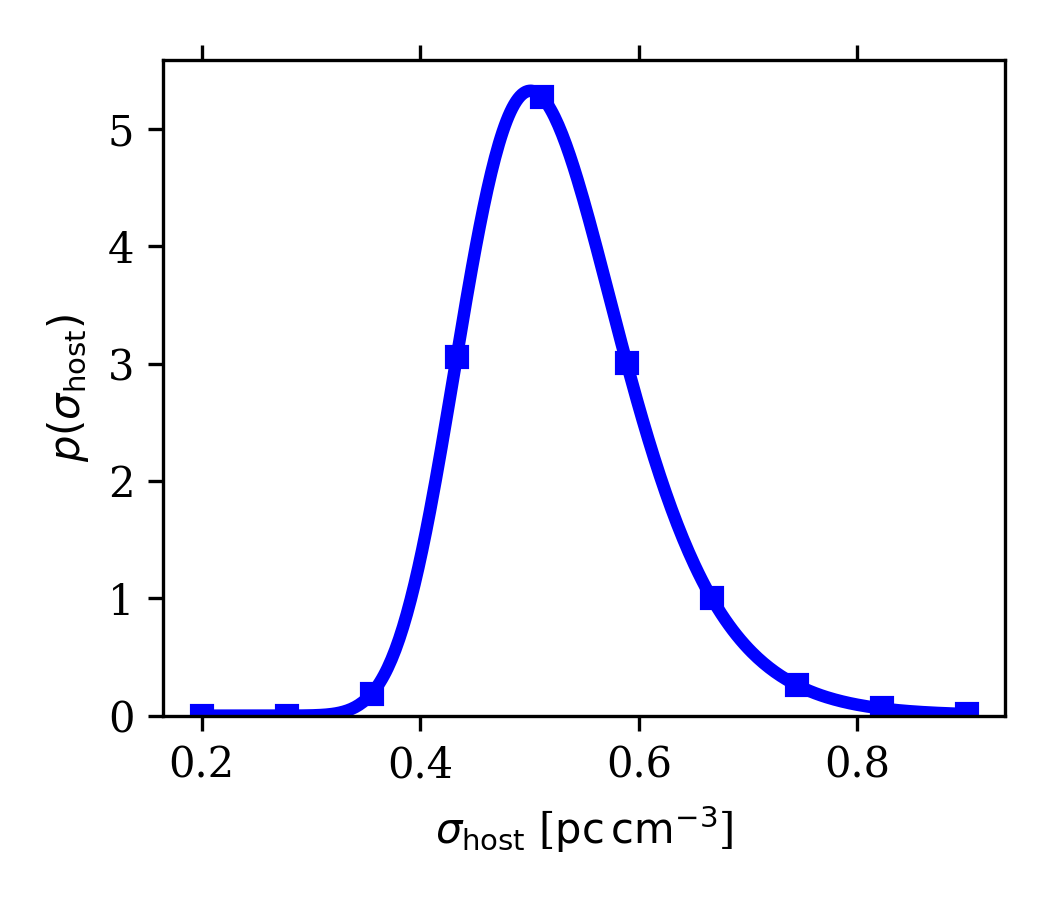}
    \includegraphics[width=2.5in]{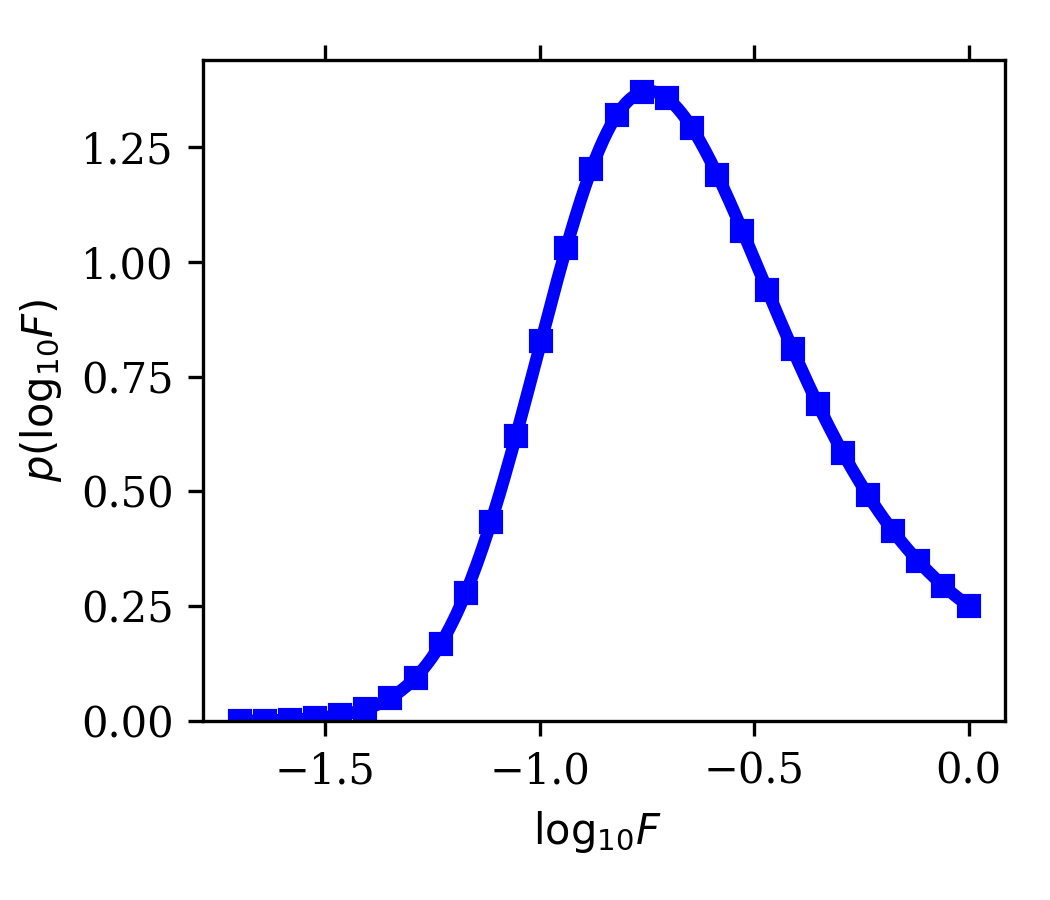}
    \caption{The calculated 1D likelihood functions using \NFRB FRBs (\NFRBz FRBs with redshifts). $F$ is measured to be $\log_{10} F = \FnoPrior$ with no priors on $H_0$. There is a loss of constraining power on $H_0$ compared to the measurement by \citet{j22b} when allowing the $F$ parameter to vary. }
    \label{fig:realcube_pdfs}
\end{figure*}

In Figure \ref{fig:realcube_pdfs}, we present the 1D PDFs of each parameter determined from the \NFRB FRBs collected from the ASKAP and Parkes 
Radio Telescopes. In comparison to \citet{j22b}, there is a significant loss of constraining power on $H_0$ by including $F$ as a free parameter. We measure a Hubble constant to be \Hubble\, which is 1.5 times more uncertain than the $H_0$ measurement in \citet{j22b}. We attribute the uncertainty to the degeneracy between $H_0$ and $F$ as indicated by the strong anti-correlation in Figures~\ref{fig:pzdm_degenerate} and \ref{fig:realcube_2D_corr}. 

\begin{figure}
    \centering
    \includegraphics[width=3.5in]{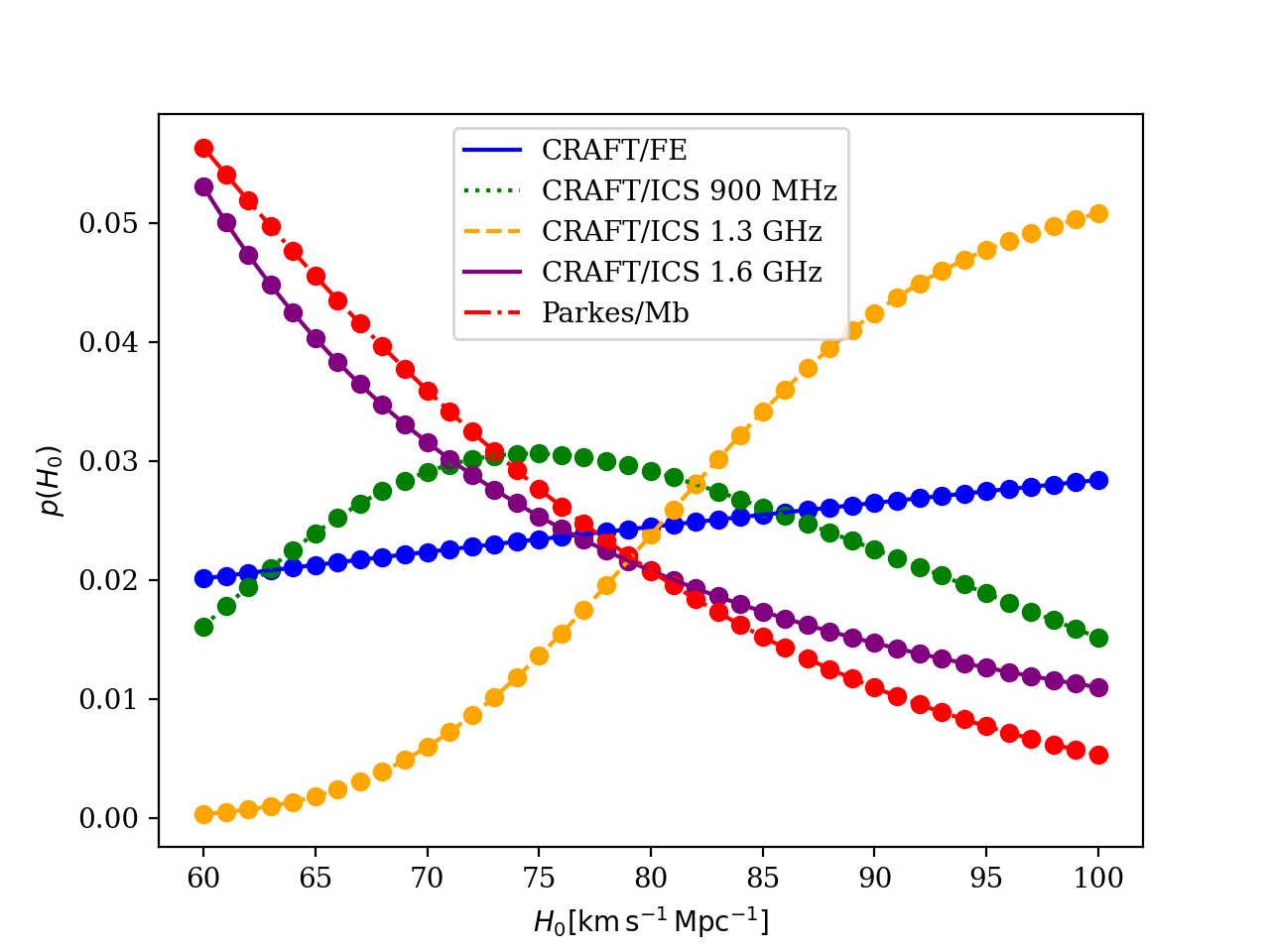}
    \includegraphics[width=3.5in]{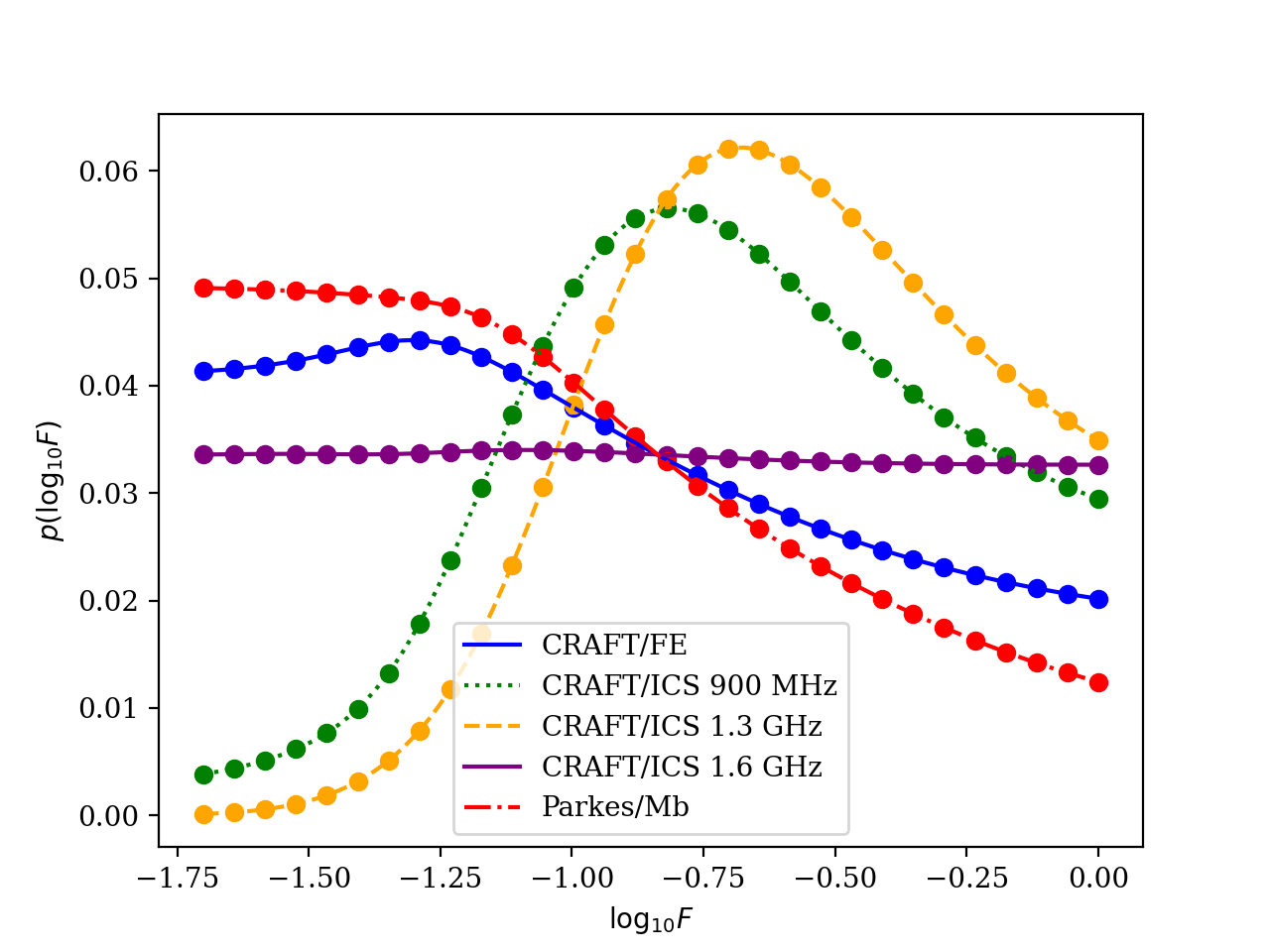}
    \caption{\textbf{Upper panel}: 1D likelihood functions of $H_0$ based on different FRB surveys used in this work. Compared to the survey contribution constraints \citep[see Figure 7 in ][]{j22b}, the constraining power of each survey is diminished. \textbf{Lower panel}: Same as upper panel for likelihood functions of $\log_{10} F$. The CRAFT/ICS 900 MHz and 1.3 GHz surveys provide the most constraining power on $F$ and $H_0$ as they contain more redshifts than the other surveys.}
    \label{fig:survey_cont}
\end{figure}

\begin{figure*}
    \centering
    \includegraphics[width=3.3in]{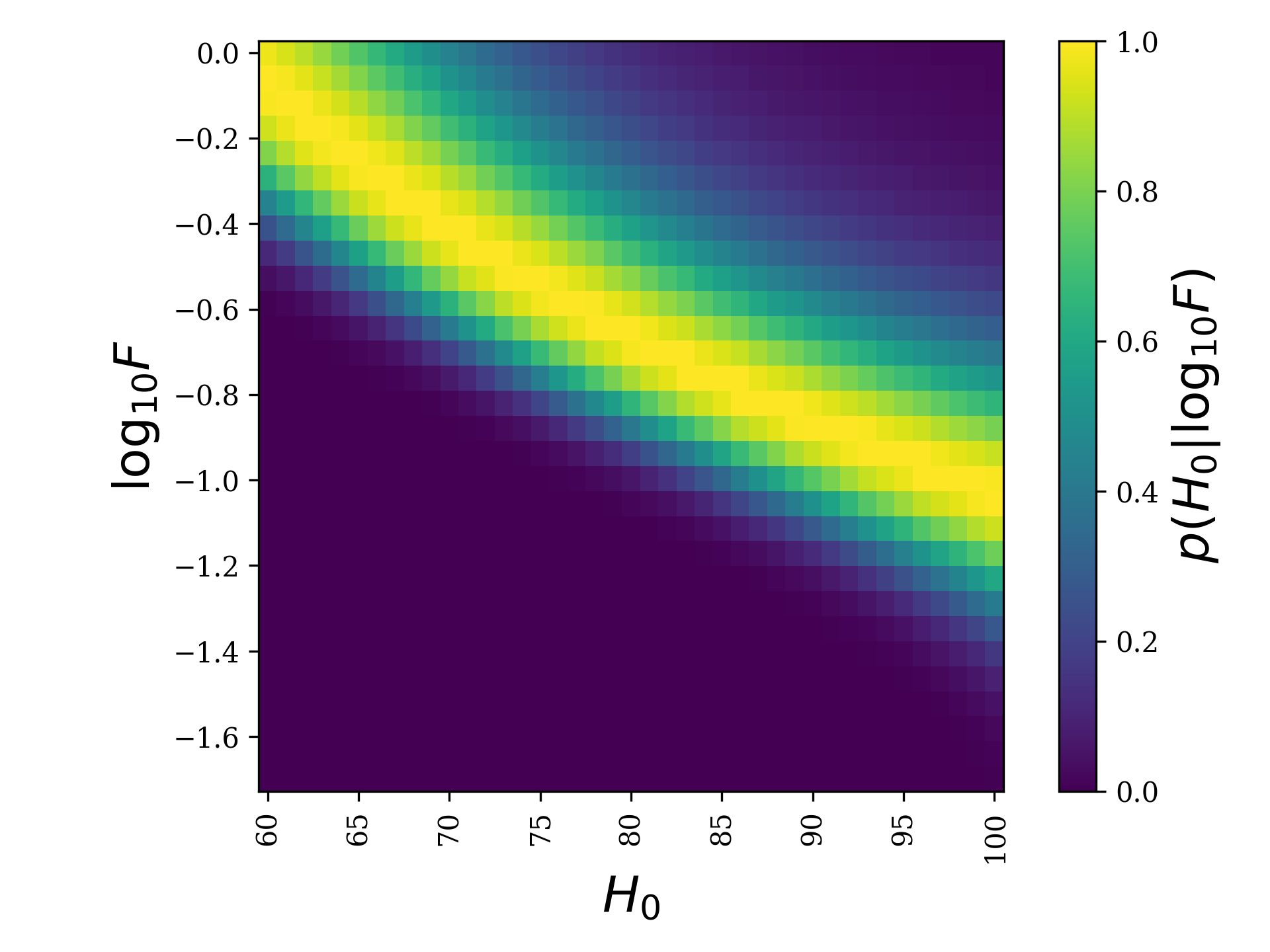}
    \includegraphics[width=3.3in]{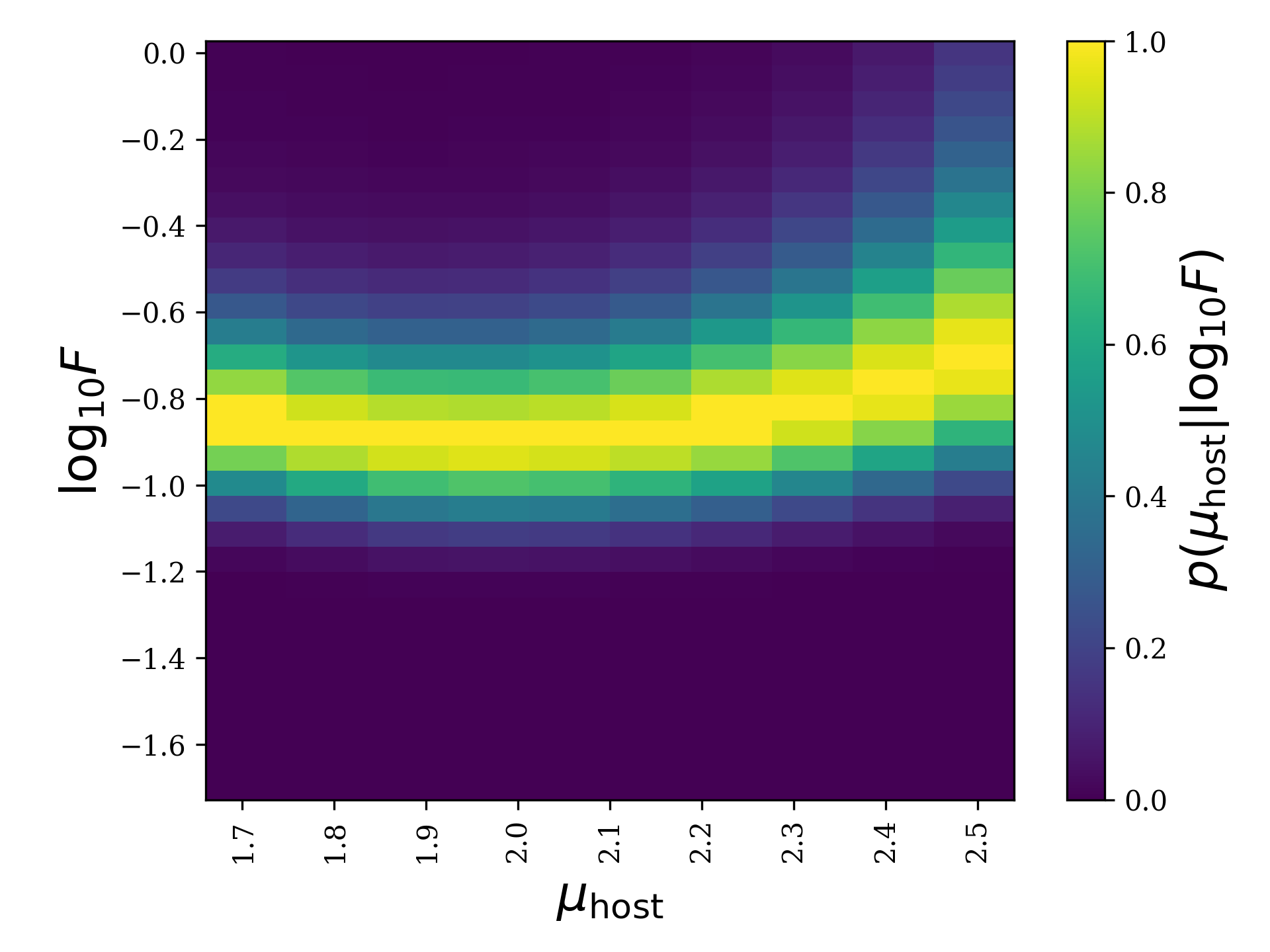} \\
    \includegraphics[width=3.3in]{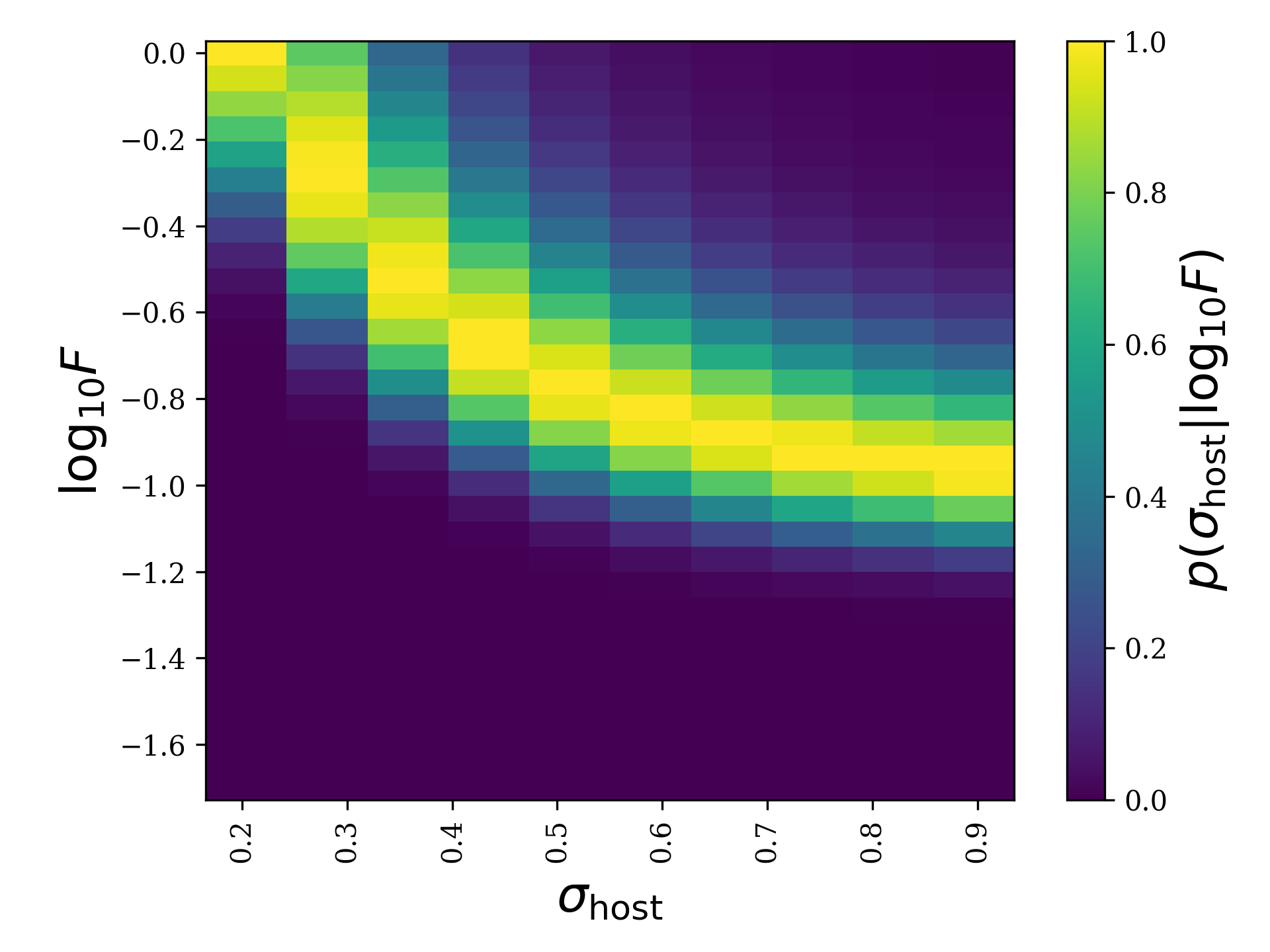}
    \caption{The 2D likelihood functions for each parameter compared against $\log_{10} F$ derived from \NFRB FRBs (\NFRBz FRBs with redshifts). There is a strong anti-correlation between $F$ and $H_0$. Additionally, we observe strong correlations between $F$ and the host galaxy \dmeg\, contribution ($\lhost, \lsigma$).}
    \label{fig:realcube_2D_corr}
\end{figure*}

To understand how our survey data at different frequencies contribute to the constraining power on $H_0$ and $F$, we replicate the survey contribution determination from \citet{j22b} which provides the 1D parameter likelihood across different FRB surveys with the Murriyang (Parkes) and Australian Square Kilometre Array (ASKAP). Figure \ref{fig:survey_cont} shows the 1D PDFs of $H_0$ and $\log_{10} F$ across the different surveys used in this analysis. We observe that the CRAFT 1.3 GHz and 900 MHz surveys tend to have stronger constraining power as they contain more FRBs with measured distances (10 and 7 redshifts respectively) than the rest of the surveys.

In Figure \ref{fig:realcube_2D_corr}, we present the 2D likelihoods of each parameter against $F$.  We observe correlations between $F$ and the FRB host galaxy parameters ($\lhost$, $\lsigma$), which we expect to be degenerate given that they both influence the variance of $\dmacosmic$. 
As expected from Figure \ref{fig:pzdm_degenerate}, the degeneracy in the lower bound (or cliff) of the \zdmeg\, distribution results in the strong anti-correlation of $H_0$ and $F$, resulting in a loss of constraining power on $H_0$ when allowing $F$ to vary. 

Our initial simultaneous fit does not implement any priors on the model parameters. As motivated by the $H_0$-$F$ degeneracy in Figure \ref{fig:pzdm_degenerate}, we determine the 1D likelihoods of $\log_{10} F$ by limiting our grid to different values of $H_0$. We consider a uniform prior on $H_0$ between 67.4 and 73.04 $\hunits$—the lower bound is motivated by the $H_0$ constraint from \citet{planck18}, and the upper bound is motivated by cosmological constraints using type-1a supernovae (SNe) from \citet{riess2021}.

\begin{figure}
    \centering
    \includegraphics[width=3.3in]{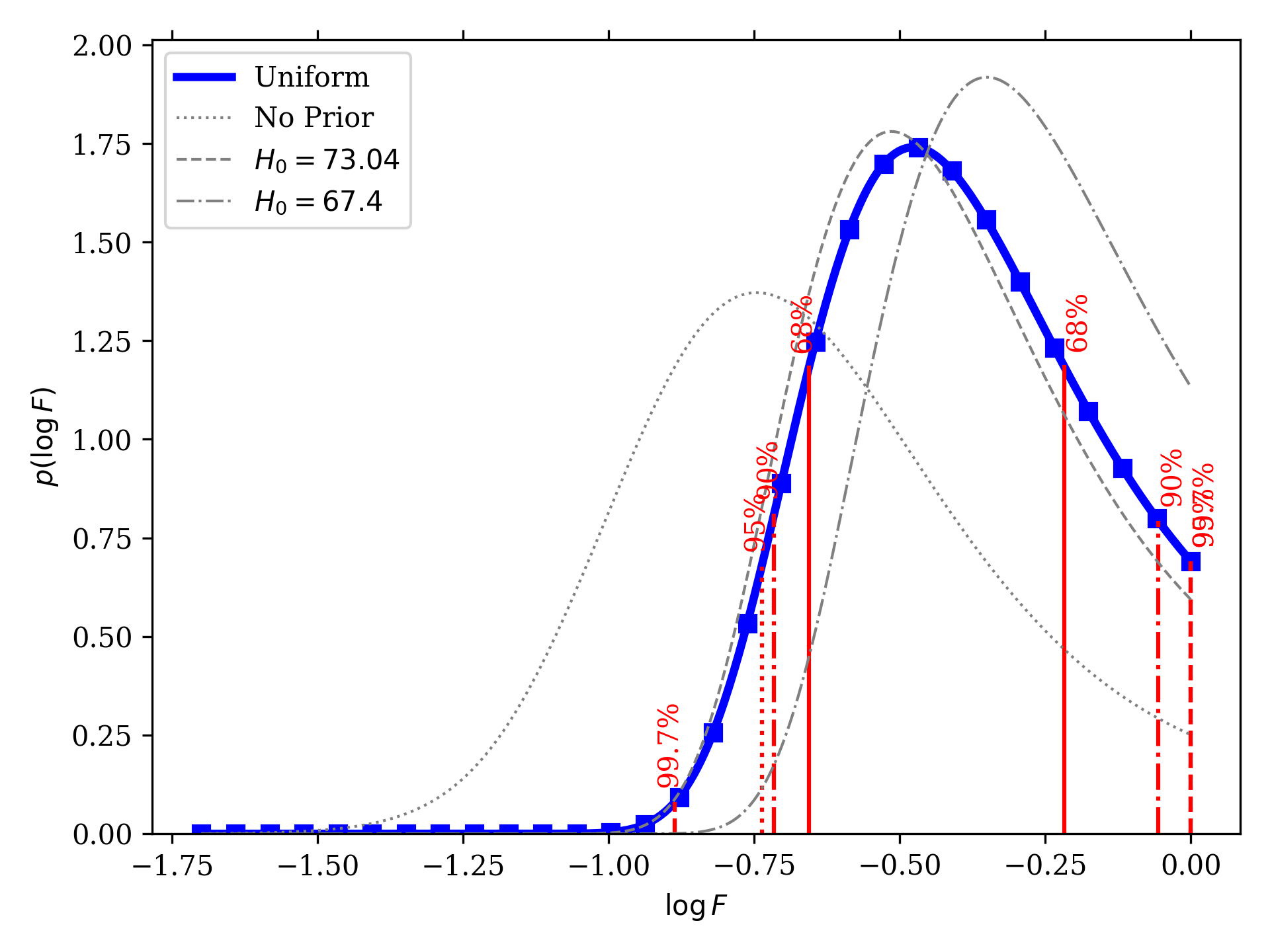}
    \caption{1D survey likelihoods of $\log_{10} F$ assuming different priors on $H_0$. The \textbf{dotted gray line} is the original 1D likelihood without any priors on $H_0$. The \textbf{blue line} is the likelihood which adopts a uniform prior on $H_0 \in [67.4, 73.04]$. The \textbf{dashed gray line} is the likelihood adopting $H_0 = 67.0 \, {\rm km s^{-1} Mpc^{-1}}$. The \textbf{dash-dotted gray line} is the likelihood adopting $H_0 = 73.0 \, {\rm km s^{-1} Mpc^{-1}}$. Adopting a prior on $H_0$ or fixing the values of $H_0$ greatly improves the constraint on $F$.}
    \label{fig:realcube_forecast}
\end{figure}

In Figure \ref{fig:realcube_forecast}, we present the 1D likelihood of the fluctuation parameter assuming different priors on $H_0$. Assuming a uniform prior between the CMB and SNe-derived values of $H_0$, we measure the fluctuation parameter to be $\log_{10} F = \FwPrior$ within $1 \sigma$ ($\log_{10} F = \Flower$ with 99.7\% confidence). We present all measurements of the $F$ parameter with different priors on $H_0$ in Table \ref{tab:logF_measurements}.

\begin{deluxetable*}{ccccc}
\tablewidth{24pt}
\tablecaption{Measurements of the $F$ Parameter \label{tab:logF_measurements}}
\tabletypesize{\normalsize}
\tablehead{
\colhead{Survey} & 
\colhead{No Prior} &
\colhead{Uniform $H_0$ Prior} &
\colhead{CMB $H_0$} &
\colhead{SNe $H_0$}
}
\startdata 
Observed & \FnoPrior & \FwPrior & \FCMB & \FSNe \\
Synthetic & \fctFnoPrior & \fctFwPrior & \fctFCMB & \fctFSNe \\
\hline
\enddata
\tablecomments{This table lists the measurements of the $F$ parameter from the observational FRB survey (76 localized FRBs with 16 redshifts) and the synthetic CRACO survey (100 localized FRBs; all with redshifts). The measurements are presented without a prior, a uniform prior between the CMB ($H_0 = 67.0 \, {\rm km\, s^{-1} \,Mpc^{-1}}$) and SNe ($H_0 = 73.0 \, {\rm km\, s^{-1} \,Mpc^{-1}}$) with their respective Gaussian errors on each side, and fixing $H_0$ to the CMB or SNe estimates.}

\end{deluxetable*}

\subsection{Parameter Likelihoods from Synthetic Surveys}
\label{ll_forecast}
\begin{figure*}
    \centering
    \includegraphics[width=2.5in]{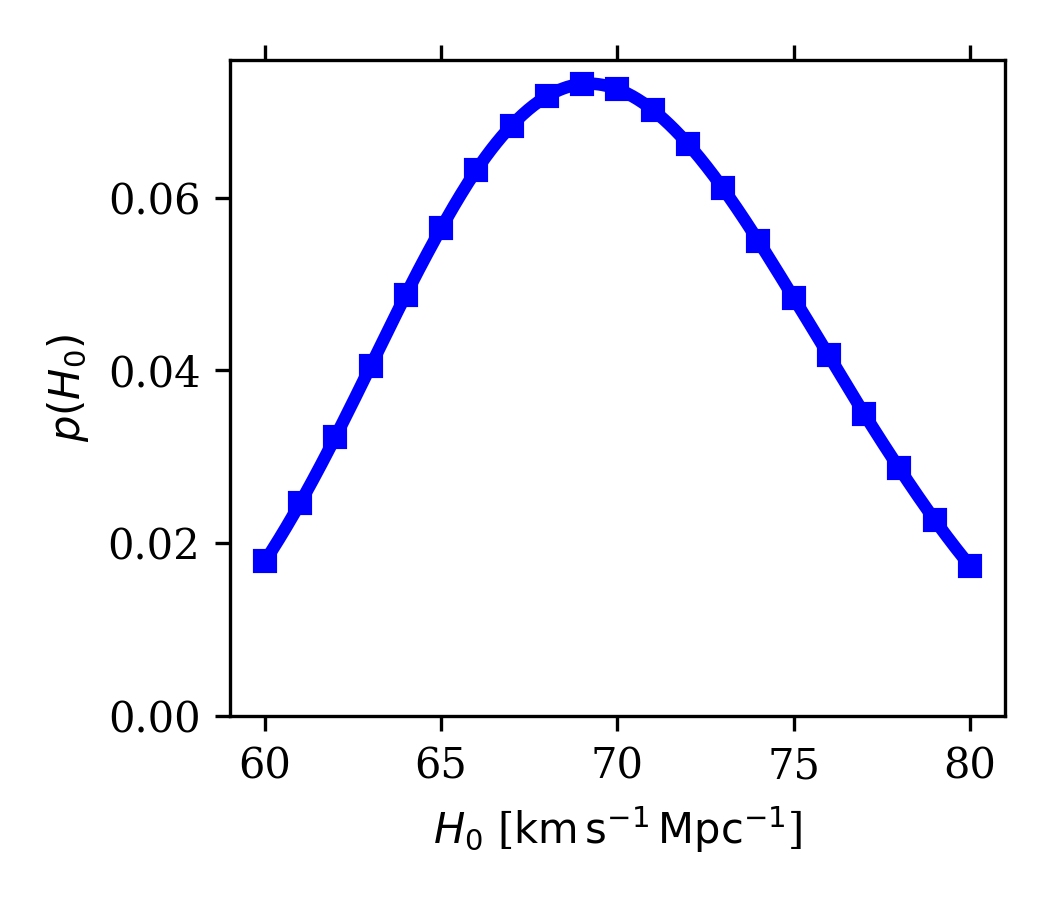}
    \includegraphics[width=2.5in]{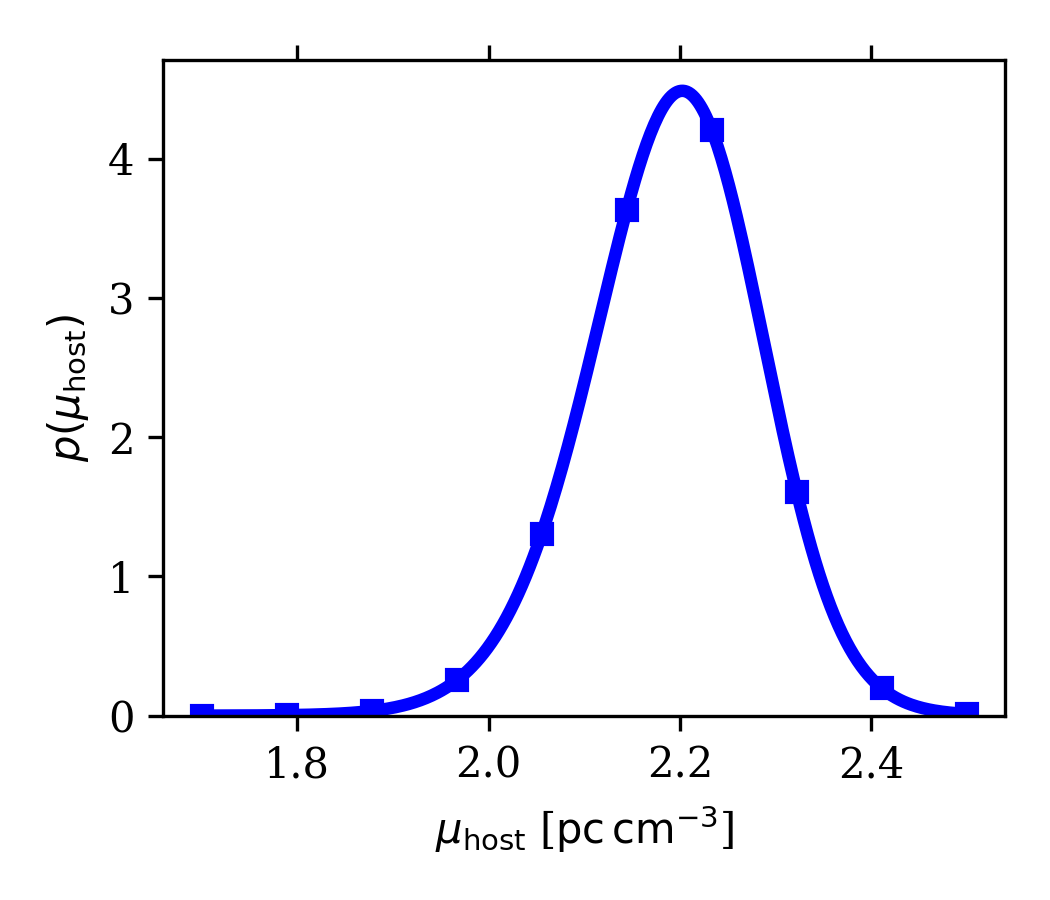}
    \includegraphics[width=2.5in]{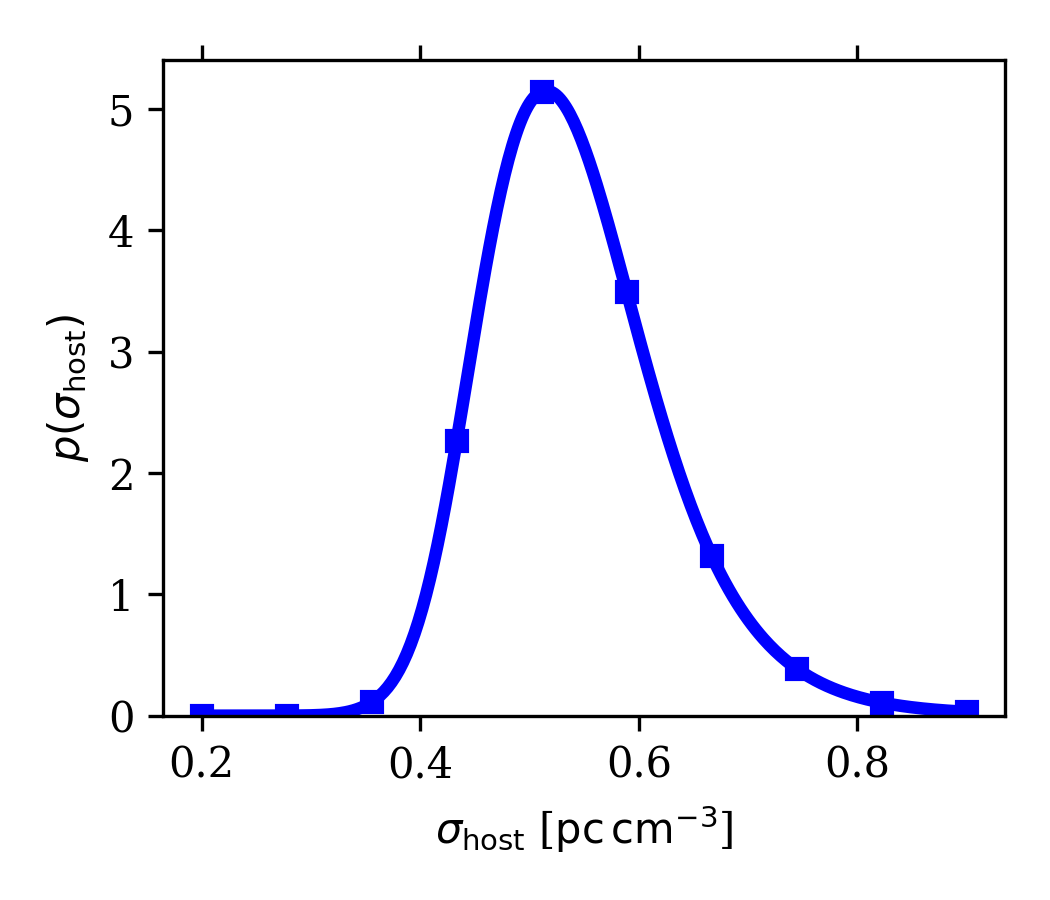}
    \includegraphics[width=2.5in]{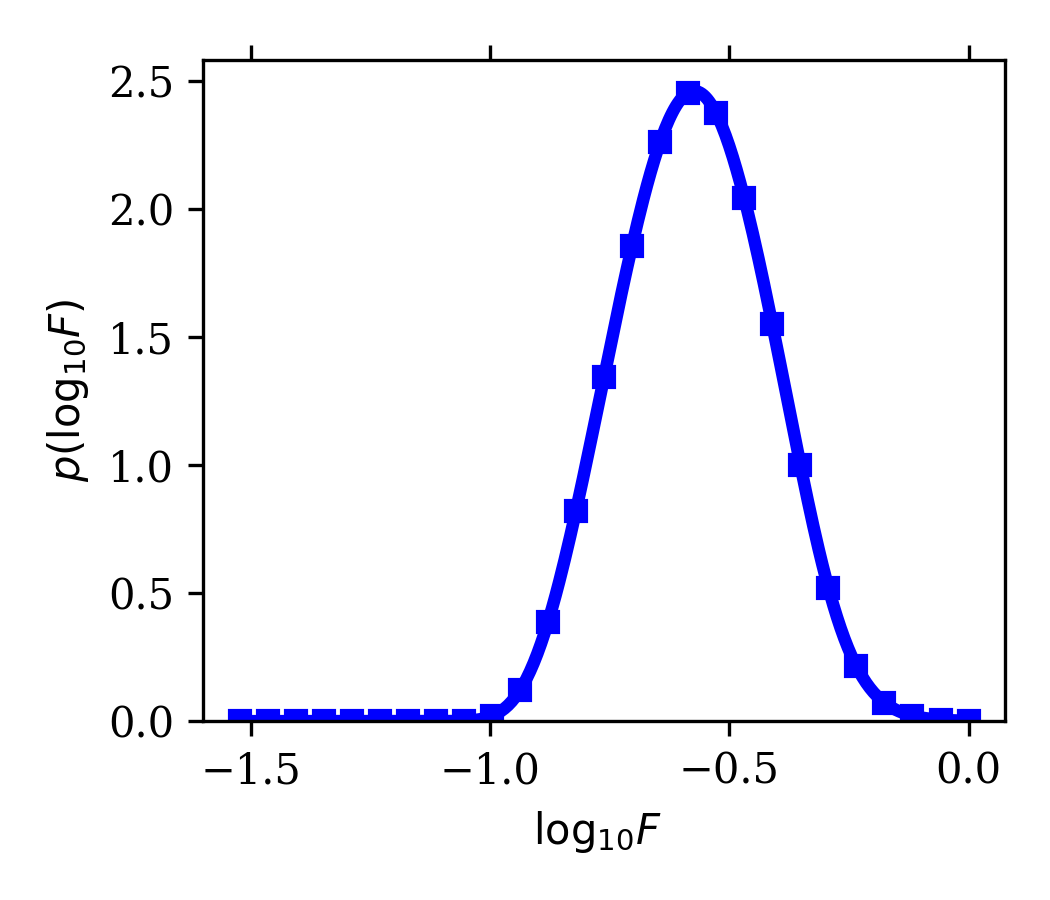}
    \caption{The 1D likelihood functions for each parameter using 100 synthetic FRBs. The constraint on $F$ is enhanced as the uncertainty due to sample size is reduced. Similarly, the constraint on $H_0$ is improved compared to the observational fit with only 21 redshifts.}
    \label{fig:fullcube_pdfs}
\end{figure*}


We use a synthetic sample of 100 localized CRACO FRBs to investigate the improvement in constraining power on both $F$ and $H_0$. In Figure \ref{fig:fullcube_pdfs}, we present the PDFs of each parameter in the grid. We observe that the constraint on $H_0$ has significantly improved by a factor of 1.7 and is more Gaussian than the previous run with \fctH. Assuming a survey of 100 localized FRBs, the best measurement we can make on $H_0$ if we adopt a Gaussian prior on $\log_{10} F$ (assuming $1\sigma$ corresponds to a 20\% error in the measurement) is $\fctHwPrior \, \hunits$ (see Table \ref{tab:H0_measurements}).

\begin{figure}
    \centering
    \includegraphics[width=3in]{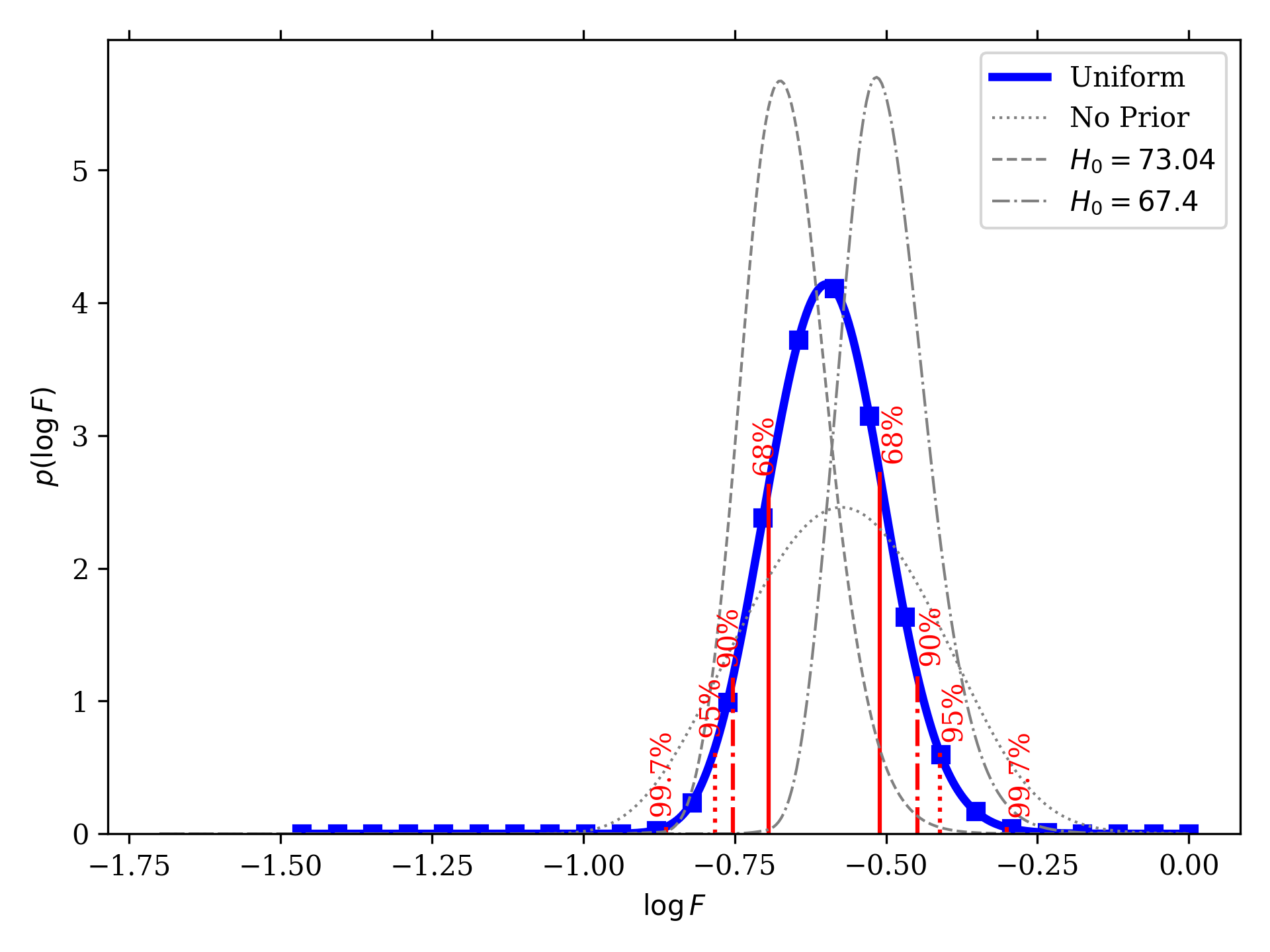}
    \caption{Same as Figure \ref{fig:realcube_forecast} but based on fits to the synthetic FRB sample. Assuming a uniform prior between CMB and SNe values of $H_0$ equipped with their associated Gaussian errors on both sides, we find $\log_{10} F = \fctFwPrior$. Fixing the value of $H_0$ also greatly enhances the constraint on $F$ by a factor of $>1.5$.}
    \label{fig:fullcube_forecast}
\end{figure}

In Figure \ref{fig:fullcube_forecast} we show posterior estimates for $\log_{10} F$ using different priors (see Table \ref{tab:logF_measurements}). Using the uniform prior, we obtain a forecast on the fluctuation parameter of $\log_{10} F = -0.60_{-0.18}^{+0.19}$ within $2 \sigma$. We note that when compared to Figure \ref{fig:realcube_pdfs}, there is a definitive upper limit on the fluctuation parameter rather than only a lower limit. Incorporating the uniform prior enhances the constraint on $F$ by a factor of $\sim 1.5$, and fixing the value of $H_0$ can increase the constraint by a factor of $2.5$. 

\begin{deluxetable}{ccc}
\tablewidth{36pt}
\tablecaption{Measurements of $H_0$ \label{tab:H0_measurements}}
\tabletypesize{\normalsize}
\tablehead{
\colhead{Survey} & 
\colhead{No Prior} &
\colhead{Gaussian Prior}
}
\startdata 
Observed & \Hubble & \HwPrior  \\
Synthetic & \fctH & \fctHwPrior \\
\hline
\enddata
\tablecomments{This table lists the measurements of $H_0$ from the observational FRB survey (76 localized FRBs with 16 redshifts) and the synthetic CRACO survey (100 localized FRBs; all with redshifts). The measurements are presented without a prior on $F$ and a Gaussian prior on $F$ centered at $\log_{10} F \simeq -0.49$ with $\sigma \simeq 0.1$ (20\% error on $F$).}
\end{deluxetable}

\section{Discussion} \label{sec:disc}


\subsection{Measurement of the fluctuation parameter}

Our principle result from the population analysis
of \NFRB~FRBs (\NFRBz with redshifts) is a lower limit on $F$ which is $\log_{10} F = \FwPrior$ (\Flower\, at 99.7\% confidence). This measurement is motivated by \citet{j22b}, where they noted that for future localization of FRBs beyond $z\gtrsim 1$, $F$ may need to be fitted explicitly. We note that this observation is only made when adopting a prior between the CMB and SNe values of $H_0$.

\subsection{Fluctuation Parameter Degeneracies}
Our findings indicate a strong degeneracy between the Hubble constant $H_0$ and the fluctuation parameter $F$ when simultaneously fitting both within the \textsc{z-DM} modeling framework adopted by \citet{j22a} which uses $F=0.32$ which falls within the accepted range of our measurement.

Aside from the degeneracy between $F$ and $H_0$, we would like to call attention to the possible degeneracy between $F$ and $\sigma_8$--the RMS amplitude of the matter density field when smoothed with an $8h^{-1}$ Mpc filter. In the case of w feedback ($F \rightarrow 1$), more mass would be concentrated within cosmic filaments, increasing the variance of a fixed-mass filter (i.e., $\sigma_8$). We expect these two parameters to be inversely coupled. A preliminary analysis varying $\sigma_8$ in the CAMELS IllustrisTNG cosmological simulations 
does show a positive correlation between $F$ and $\sigma_8$ (Medlock et al. in prep.).

\subsection{Forecasting enhanced constraints on $F$}
Using a sample of 100 synthetic FRBs (see Figure \ref{fig:fullcube_forecast}), we are able to constrain both upper and lower limits on the fluctuation parameter out to $3 \sigma$. Since we are only able to effectively constrain a lower limit on $F$, we compare the lower-sided half-maximum widths. We find the left-sided half-maximum width of the synthetic distribution is \textit{half} the width of the current measured distribution. We expect this constraint to only improve with more localizations, which will be easily facilitated with next-generation all-sky radio observatories.

Additionally, it is of interest to see how this method compares to other ways of measuring the baryon distribution in the IGM. For example, an alternative method to constrain AGN and stellar feedback focuses on small-scale deviations in the matter power spectrum \citep{vdaalen20}. As baryonic feedback significantly influences the mass distribution at smaller scales (higher $k$), probes of the gas density at those scales (thermal Sunyaev-Zel'dovich effect) can measure the intergalactic baryon distribution \citep{pandey23}.

\subsection{Comparing with Fluctuation Parameter in IllustrisTNG}

In a work by \citet{zhang21} to highlight the utility of FRBs in probing the IGM, they generated thousands of FRB sightlines in IllustrisTNG and fitted the observed extragalactic DM excess $p_{\rm cosmic}(\Delta)$. They provide the fitted parameters as well as the dispersion in the \zdmeg\, distribution $\sigma_{\rm DM}$. We convert these values into the fluctuation parameter $F$ by assuming $\sigma_{\rm DM} = Fz^{-0.5}$. 

\begin{figure}
    \centering
    \includegraphics[width=.45\textwidth]{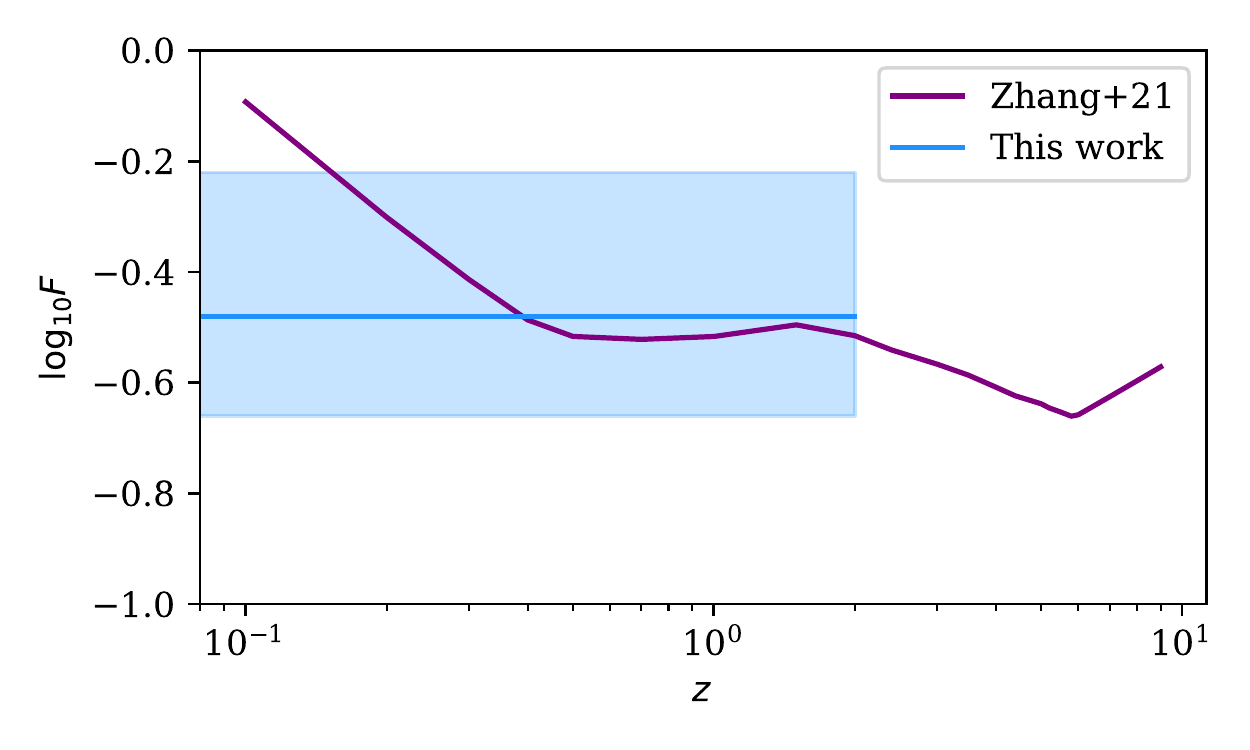}
    \caption{Fluctuation parameters derived from this work, the fiducial value from \citet{j22b}, and the IllustrisTNG values from \citet{zhang21}. Our measurement on $F$ agrees with the simulated $F$ parameter between $0.4 < z < 2$.}
    \label{fig:illustris}
\end{figure}

In Figure \ref{fig:illustris}, we present these derived $\log_{10} F$ values as a function of redshift compared to our measured values. Between $0.4 < z < 2$, our measurements are
in fine agreement. However, we observe that the fluctuation parameter in Illustris appears to be higher at $z \lesssim 0.4$ and lower when $z > 2$.

From the redshift-dependent $\dmigm$ distributions derived from IllustrisTNG (figure 2 from \citet{zhang21}), distributions between $0.1 < z < 0.4$ are wider and the modes of each distribution are spread further apart. This may explain why the IllustrisTNG fluctuation parameter is higher than our measurement as the $\dmigm$ distribution functions have larger variance at those redshifts.


To make a proper comparison between our work and \citet{zhang21}, it may be necessary to introduce a free parameter for
the redshift evolution of $\sigma_{\rm DM}$ 
instead of simply fixing the redshift exponent to $-1/2$ 
(Equation \ref{eq:sigma_dm}). 

\section{Conclusions} \label{sec:conc}

In this work, we have implemented 
variance in \dmcosmic\ as a free parameter in a 
forward model of the \zdmeg\, distribution of FRBs. With this adapted model and a survey of \NFRB ASKAP and Parkes FRBs, we constrain a value for the fluctuation parameter, explore degeneracies within the model, and generate a forecast of the constraint on the fluctuation parameter with a synthetic survey of 100 localized FRBs. The conclusions we draw from this analysis are:

\begin{itemize}

\item Incorporating survey data of \NFRB (\NFRBz with redshifts) FRBs yields
a firm lower limit on $F$. We place the lower limit on $F$ as measured by the survey sample to be $\log_{10} F > -0.89$ at 99.7\% confidence. The 900 MHz and 1.3 GHz surveys dominate this constraint due to their higher number of localizations to host galaxies and
their associated redshifts.

\item Forward modeling the FRB data from Parkes and ASKAP, the fluctuation parameter is degenerate with the Hubble constant $H_0$.


\item We forecast that 100 localized FRBs are sufficient to constrain both an upper and lower limit on the fluctuation parameter. With the greater count of localizations, the half-maximum width of the distribution decreases by $\approx 50\%$.

\item Extrapolation of the fluctuation parameter from IllustrisTNG shows agreement between $0.4 < z < 2.0$. \citet{zhang21} measure a higher fluctuation parameter at low redshift ($z < 0.4$) and a lower fluctuation parameter beyond $z > 2$. The former result is likely to be an effect of the rapidly evolving $\dmigm$ distribution at low redshift. 

\end{itemize}

Next-generation radio observatories will significantly improve the constraint on the fluctuation parameter. For example, the Deep Synoptic Array 2000 (DSA-2000) is expected to localize on the order of 10,000 FRBs each year --- enough FRBs to sufficiently characterize the baryonic contents of the IGM \citep{hallinan19, ravi19}.

Additionally, the FRB Line-of-sight Ionization Measurement From Lightcone AAOmega Mapping (FLIMFLAM) survey is an upcoming spectroscopic survey that seeks to map the intervening cosmic structures and diffuse cosmic baryons in front of localized FRBs \citep{lee22}. These FRB foreground data taken in the Southern hemisphere will be used in conjunction with ASKAP FRB measurements to improve the constraints on the intergalactic baryon distribution \citep{lee22}.

These expansions in FRB surveys with localizations are expected to greatly improve the constraints on the fluctuation parameter. With these improved constraints on $F$, one may leverage this
novel observable for investigating feedback and cosmological prescriptions in simulations.

Combining the $F$ parameter with other observables like the thermal SZ effect that trace the intergalactic baryon distribution, there is ample opportunity to better inform subgrid feedback models \citep{munoz18, pandey23}.

\section*{Acknowledgements}
J.B. acknowledges support from the University of California Santa Cruz under the Lamat REU program, funded by NSF grant AST-1852393, and the Yale Science Technology and Research Scholars Fellowship funded by the Yale College Dean's Office. Authors A.G.M. and J.X.P., as members of the Fast and Fortunate for FRB Follow-up team, acknowledge support from NSF grants AST-1911140, AST-1910471 and AST-2206490.
The authors acknowledge the use of the Nautilus
cloud computing system which is supported
by the following
US National Science Foundation (NSF) awards:
CNS-1456638, CNS-1730158, CNS-2100237, CNS-2120019, ACI-1540112, ACI-1541349, OAC-1826967, OAC-2112167.

CWJ and MG  acknowledge support by the Australian Government through the Australian Research Council's Discovery Projects funding scheme (project DP210102103).

RMS and ATD acknowledge support through Australian Research Council Future Fellowship FT190100155 and Discovery Project DP220102305.

The Australian SKA Pathfinder is part of the Australia Telescope National Facility (https://ror.org/05qajvd42) which is managed by CSIRO. Operation of ASKAP is funded by the Australian Government with support from the National Collaborative Research Infrastructure Strategy. ASKAP uses the resources of the Pawsey Supercomputing Centre. Establishment of ASKAP, the Murchison Radio-astronomy Observatory and the Pawsey Supercomputing Centre are initiatives of the Australian Government, with support from the Government of Western Australia and the Science and Industry Endowment Fund. We acknowledge the Wajarri Yamatji people as the traditional owners of the Observatory site.

This research is based on observations collected at the European Southern Observatory under ESO programmes 0102.A-0450(A), 0103.A-0101(A), 0103.A-0101(B), 105.204W.001, 105.204W.002, 105.204W.003, 105.204W.004, 108.21ZF.001, 108.21ZF.002, 108.21ZF.005, 108.21ZF.006, and 108.21ZF.009.

\bibliography{frb_f}

\end{document}